\documentclass[referee]{raa}
\usepackage{graphicx,times}
\usepackage{natbib}
\usepackage{amssymb,amsmath}
\bibpunct{(}{)}{;}{a}{}{,}
\newcommand{\grad}{\mathop{\rm grad}\nolimits}

\usepackage[a4paper=true,dvipdfm=true,pagebackref=true]{hyperref}
\hypersetup{colorlinks = true, linkcolor = green, anchorcolor = red, citecolor = blue, filecolor = red, pagecolor = red, urlcolor = red}

\begin{document}

   \title{Modeling the vertical distribution of the Milky Way's flat subsystem objects}

 \volnopage{ {\bf 20XX} Vol.\ {\bf X} No. {\bf XX}, 000--000}
   \setcounter{page}{1}

   \author{Igor' I. Nikiforov, Vadim A. Usik and Angelina V. Veselova \inst{}} 



   \institute{Saint Petersburg State University, Universitetskij Prospekt 28, Staryj
Peterhof, Saint Petersburg 198504, Russia; {\it i.nikiforov@spbu.ru (IIN)}
\\ 
\vs \no {\small Received 20XX Month Day; accepted 20XX Month Day}
}

\abstract{
This paper is an initial stage of consideration of the general problem of joint modeling of the vertical structure of a Galactic
flat subsystem and the average surface of the disk of the Galaxy, taking into account the natural and measurement dispersions. We
approximate the average surface of the Galactic disk in the region covered by the data with a general (polynomial) model and
determine its parameters by minimizing the squared deviations of objects along the normal to the model surface. The smoothness of
the model, i.e., its order~$n$, is optimized. An outlier elimination algorithm is applied. The developed method allows us to
simultaneously identify significant details of the Galactic warping and estimate the offset~$z_{\sun}$ of the Sun relative to the
average (in general, non-flat) surface of the Galactic disk and the vertical scale of the object system under consideration for an
arbitrary area of the disk covered by data. The method is applied to data on classical Cepheids (\citeauthor{Berdnikov+2000},
\citeauthor{Mel'nik+2015}). Significant local extremes of the average disk surface model were found based on Cepheid data: the
minimum in the first Galactic quadrant and the maximum in the second.  A well-known warp
(lowering of the disk surface) in the third quadrant has been confirmed. The optimal order
of the model describing all these warping details was found to be $n_\text{o}=4$. The
local (for a small neighborhood of the Sun, $n_\text{o}=0$) estimate of $z_{\sun} = 28.1
\pm
\left.6.1\right|_{\text{stat.}}\left.{}\pm1.3\right|_{\text{cal.}}$\,pc
is close to the non-local (taking into account warping, $n_\text{o}=4$) $z_{\sun} = 27.1
\pm \left.8.8\right|_{\text{stat.}}\left.{}^{+1.3}_{-1.2}\right|_{\text{cal.}}$\,pc
(statistical and calibration uncertainties are indicated), which suggests that the
proposed modeling method eliminates the influence of warping on the $z_{\sun}$ estimate.
However, the non-local estimate of the vertical standard deviation of Cepheids
$\sigma_{\rho} = 132.0 \pm
\left.3.7\right|_{\text{stat.}}\left.{}^{+6.3}_{-5.9}\right|_{\text{cal.}}$\,pc differs
significantly from the local $\sigma_{\rho} = \left.76.5 \pm
4.4\right|_{\text{stat.}}\left.{}^{+3.6}_{-3.4}\right|_{\text{cal.}}$\, pc, which means
the need to introduce more complex models for the vertical distribution outside the Sun's
vicinity.
%by \citeauthor{Berdnikov+2000} in the version by \citet{Mel'nik+2015}
\keywords{Galaxy: disk  --- Galaxy: structure --- Galaxy: fundamental parameters ---
methods: data analysis
}
}

   \authorrunning{Nikiforov et al.}            %author_head in even pages
   \titlerunning{Modeling the vertical distribution of disk objects}  % title_head in odd pages
   \maketitle

%________________________________________________ sections below
% 
\section{Introduction}           %% first-level sections will be auto-capitalized
\label{sect:intro}

The vertical distribution of objects in various Galactic subsystems contains valuable
information about the origin, evolution and dynamics of our Galaxy, so determining the
characteristics of this distribution is an important task of Galactic Astronomy.
The study of the vertical distribution may include consideration of many phenomena, but
one of them should be taken into account necessarily---this is the offset $z_{\sun}$ of
the Sun relative to the plane of the Milky Way's disk towards the North Galactic Pole. 
Therefore, in the simplest case, modeling of the vertical distribution is reduced to
determining the value of~$z_{\sun}$ and some dispersion parameter (standard deviation,
scale height, etc.) that characterizes the scattering of subsystem objects relative to the
average plane of the Galactic disk (usually relative to the midplane of this subsystem).
The first estimate of $z_{\sun} = 13.5 \pm 1.7 \,\text{pc}$ was obtained by
\citet{vanTulder1942} from the analysis of nearby stars. Subsequently, in many papers, the
solar offset was determined by different methods for various objects and Galactic
subsystems. \citet{Bland-Hawthorn+2016} adopted as the best (local) estimate the result of
\citet{Juric+2008} from the complete SDSS photometric survey, $z_{\sun} = 25 \pm
5\,\text{pc}$, which covers many other estimates.

However, the solar offset relative to the differently defined midplane of the disk does
not seem to be described by a single value of~$z_{\sun}$. For example,
\citet{Bobylev+Bajkova2016_1} obtained significantly different results for reference
objects (tracers) of different types: $z_{\sun} = 5.7 \pm 0.5$ pc for a sample of methanol masers,
$z_{\sun} = 7.6 \pm 0.4$ pc for data on H\,II regions and $z_{\sun} = 10.1 \pm 0.5$ pc for
data on giant molecular clouds;  at the same time, \citet{Ferguson+2017} derived
values of $z_{\sun} =14.9 \pm 0.5$\,pc for a uniform selection of SDSS K and M dwarf
stars and $z_{\sun}=15.3\pm0.4$\,pc for an expanded selection,
\citet{Buckner+Froebrich2014} found an estimate $z_{\sun} = 18.5 \pm 1.2\,\text{pc}$ for
open clusters, and \citet{Majaess+2009} obtained values of $z_{\sun} =26 \pm 3$\,pc
for Cepheids. A comparison of these and other~$z_{\sun}$ estimates obtained in various
studies (see, e.g., summaries in
\citealt{Yao+2017,Skowron+2019}) shows that the differences between these estimates
cannot be explained only by statistical errors, with some estimates varing significantly,
even for objects of the same type (e.g., for open clusters and Cepheids). This shows that
the discrepancies reflect not only the possible objective difference in the values of
$z_{\sun}$ between different types of objects (subsystems of the Galaxy), but also other
factors in the problem.

%However, the solar offset relative to the differently defined mid plane of the disk seems
%to be a more complex phenomenon that is not described by a single $z_{\sun}$ value.

% from data of the Sloan Digital Sky Survey.

%\citet{Skowron+2019}...
% because, for example, Cepheids with different ages trace significantly different parts
%of the disk, the younder Cepheids are observed closer to the Galactic Center.

In addition to the $Z$-offset of the Sun, the number of already established or potential
factors affecting the results of modeling the vertical distribution of objects includes:
1)~the warp of the Galactic disk, 2)~the dependence of the values of the characteristics
of the vertical distribution on the position on the disk for the selected Galactic
subsystem (e.g., the flare of the Galactic disk), 3)~the possible (and in the case of
vertical dispersion, real) dependence of these characteristics on the type of Galactic
subsystem, 4)~the need to establish the functional type of the vertical distribution and
its possible variations with the position on the disk and with the type of subsystem, as
well as 5)~taking into account the random uncertainty of heliocentric distances,
systematically distorting the true vertical distribution. The problem in general (taking
into account all these factors) has not yet been solved. Meanwhile, different combinations
of these factors may be responsible for discrepancy of the results (in particular,
of~$z_{\sun}$ estimates) in different papers. Subjective factors can also lead to this:
the choice of the general appearance of the model of the average surface of the disk,
possible mismatch of the distance scales used in different works, the dependence of the
results of modeling on the size and configuration of the disk area under consideration
(the area covered by the data).

%Warp: general:

Despite the lack of a solution to the problem in general, some of these factors and their
combinations were considered. The most important factor is the presence of a warp of the
Milky Way's disk. The warp was noticed as soon as the observation data in the 21-cm line
of neutral hydrogen appeared for the southern hemisphere (\citealt{Burke1957,Kerr1957}).
Subsequent studies (\citealt{Oort+1958}; see, e.g., \citealt{BM98} and
\citealt{Bland-Hawthorn+2016} reviews and references therein; \citealt{Skowron+2019};
\citealt{Chrobakova+2020}, among others) have shown that a significant stellar/gas warp
begins outside the solar circle, and in the inner Galaxy the disk is very close to flat,
including on the far side of the disk (\citealt{Minniti+2021}). Various data indicate that
one part of the warped disk deviates from the plane of the inner disk towards the North
Galactic Pole, the other deviates in the opposite direction.

%Warp: exclusion of warp zone:

Not taking into account the large-scale warp (if a plane parallel to the equator of the
Galactic coordinate system $b=0^\circ$ is taken as a model of the average surface of the
Galactic disk) can significantly affect the estimates of the solar offset~$z_{\sun}$ and
the vertical scales of flat subsystems (see, e.g., the dependence of these characteristics
for planetary nebulae on the size of considered near-solar region in
\citealt{Bobylev+Bajkova2017}).
One way to avoid this is to exclude the warp zone from consideration under the assumption
that in the remaining area of the disk its average surface is flat: restrictions are
imposed on the selection of tracers, for example, by the heliocentric distances~$r$ (e.g.,
$r\lesssim 4$\,kpc in
\citealt{Bobylev+Bajkova2016_2}; $r\lesssim4.5$\,kpc in~\citealt{Bobylev+Bajkova2016_1}),
by the predicted maximum warp offsets (${<}10$ pc in \citealt{Yao+2017}),  by the
distance~$R$ to the axis of rotation of the Galaxy ($R<7.0$\,kpc in
\citealt{Reid+2019}). However, the exclusion of the warp zone requires the adoption of
a specific warp model, and it is often taken simple for this and other applications: the
disk in the inner Galaxy ($R\le R_\text{w}$) is considered undisturbed, and in the outer
one ($R> R_\text{w}$) it is usually represented by a combination of a power dependence on
$R$ and a simple trigonometric function of the azimuthal coordinate (e.g.,
\citealt{BM98,Pohl+2008,Xu+2015,Yao+2017,Romero-Gomez+2019,Cheng+2020,Mosenkov+2021}).
At the same time, to describe the warp in the outer Galaxy, in most of its morphological
studies, simple symmetric models with a limited set of parameters are used---the
radius~$R_\text{w}$ at which the disk starts bending, the phase angle of the line-of-nodes
and the maximum amplitude of the warp (see, e.g., \citealt{Romero-Gomez+2019} and references
therein).

However, the warp is clearly more complicated. Firstly, the inner part of the disk is not
perfectly flat--- there are corrugations on the scale of ${\sim}30$\,pc
(\citealt[fig. 3]{Oort+1958}; \citealt{Spicker+Feitzinger1986};
\citealt[fig.~9.22]{BM98}). $N$-body simulations of the Milky Way interacting with a satellite
similar to the Sagittarius dwarf galaxy show  that repeated satellite passes can generate
local ripples, including in the inner disk (\citealt[fig.~2]{Poggio+2020}). According to
kinematics, the onset of the warp occurs at a guiding radius {\em inside\/} the Solar
circle, $R_\text{g}\lesssim 7$\,kpc (\citealt{Schoenrich+Dehnen2018}), 
or even in the center of the Galaxy (\citealt{Li+2020}). Secondly, the outer
part of the warp is also not described by a simple model---there are manifestations of
lopsidedness of the warp and twisting of its line-of-nodes (\citealt{Romero-Gomez+2019,
Chrobakova+2020}); \citet{Xu+2015} detected an oscillating asymmetry in the SDSS
main-sequence star counts on either side of the Galactic plane in the anticenter region,
between longitudes of $110^\circ < l < 229^\circ$. In addition, the morphology and
kinematics of the warp depend on the type/age of the tracers (e.g.,
\citealt{Romero-Gomez+2019,Chrobakova+2020}). Moreover, hydrodynamic modeling of the
evolution of an ensemble of stars formed in the warp shows that only younger populations
trace the warp detected by~HI (\citealt{Khachaturyants+2021}) and that the influence of
the bending waves excited by irregular gas inflow is most strongly manifested in the young
populations (\citealt{Khachaturyants+2022}). This means that the warp model, universal for
all disk subsystems of the Galaxy, can hardly be accepted.

%Warp: from kinematics:

Kinematic manifestations of the warp also indicate its asymmetry and complexity in
general, as well as the dependence of its characteristics on the age of tracers (e.g.,
\citealt{Romero-Gomez+2019,Li+2020,Cheng+2020} and references in these works).

Based on the above, the exclusion of the warp zone as a method of eliminating biases in
the vertical distribution parameters can only give a partial (local) solution to the
problem, the accuracy of which depends on the details of the accepted warp model and on its
realism in the case of the tracers under consideration and on assumptions about the
boundaries of the warp-distorted area. All these assumptions can be sources of systematic
errors. That is why it seems important to us to abandon simplified warp models and
consider the most general analytical warp model describing all the significant structural
features of the middle surface of the disk identified by the tracers under consideration.
The method of excluding the warp zone is also unsuccessful due to the presence of the disk
flaring, which begins at $R\gtrsim R_0$, where $R_0$ is the Galactic center distance
(e.g., \citealt{Reid+2019,Mosenkov+2021}), since the dependence of the dispersion
parameter on the accepted boundaries of the area ``undisturbed'' by the warp appears.

%The disk interior to 8.9 kpc is unperturbed. Between 8.9 and 12.1 kpc from the Galactic
%center, the midplane disk is perturbed up in a sinusoidal pattern, (Yao+17)

%by imposing some kind of restriction on the sample of tracers
% according to the accepted warp model
%~$Z_\text{W}$

%Warp: from kinematics, precession:

The warp is currently being actively explored in many ways. In particular, warp precession
is actively discussed (e.g., \citealt{Cheng+2020}). However, as noted by
\citet{Poggio+2020}, the precession parameters depend on our knowledge of the shape of the
warp and its differences for different stellar populations. In addition, \citet{Chrobakova+LC2021}
even raise the question of the very existence of precession, since the application of a warp model
inconsistent with the tracers used leads to a fictitious precession.

Detailed warp models are also important both for studying the dependence of $z_{\sun}$ and
vertical dispersion characteristics on the type/age of tracers, and for identifying the
cause and dynamic nature of the warp of our Galaxy, which remain unclear (\citealt{BM98,
Poggio+2020,Khachaturyants+2021}).

Note also that in the framework of an alternative approach applied by
\citet{Mosenkov+2021}--- photometric 3D decomposition of the Milky Way taking into account
flaring and warp---the parameters of the warp disk are poorly determined, since only a 2D
map is considered, whereas for creating a reliable 3D model of the warp one needs to have
a 3D distribution of stars in the Galaxy. 

Despite the fact that the best solution would be to model the $Z$-distribution of objects
taking into account all the factors mentioned at once, due to the complexity of the
overall task we focus in this paper on the task of constructing a detailed warp model with
a minimum of assumptions. We will not consider the influence of random errors in the
distance here (the selected data catalog allows this, see Sect.~\ref{sect:Data}), as well as the disk
flaring, since without taking into account errors in distances, the flaring parameters may
turn out to be strongly biased.

%Z0:

%The more results for different subsystems are obtained the more about the disk warping
%and the value of $z_{\sun}$ can be said. Results for subsystems with different ages can
%provide information about dynamics and evolution of the Galaxy.

\section{Method}
\label{sect:method}

%(Why do we use this model? Why don't we use model of Drimmel et al. (2000), but it's used by Cheng, )

%There is a plenty of models of the Galactic warp. \citet{Yao+2017} \ldots

We will study the spatial distribution of objects in the {\em heliocentric\/} Cartesian
coordinate system, which does not require taking any value of $R_0$: $X$-axis is directed
towards the Galactic center, $Y$-axis is towards the rotation of the Galaxy, $Z$-axis is
towards the North Galactic Pole.

In order to free the warp model as much as possible from pre-accepted assumptions, we will
consider as models the $\zeta_n(X,Y)$ polynomials, each of which is a Maclaurin series
expansion in the solar neighborhood up to the $n$-th order of the average $Z$-coordinate
of the Galactic disk as a function of position on the $XY$ plane:
\begin{gather}
  \begin{split}
    \zeta_n(X, Y) &= \sum_{i = 0}^{n}{\sum_{j = 0}^{i}{z_{i - j,\,j} X^{i - j} Y^j}} \\
    &= z_{00} + z_{10} X + z_{01} Y +  z_{20} X^2 + z_{11} X Y + z_{02} Y^2 +\:\ldots\:+ z_{n0}X^n +\:\ldots\:+ z_{0n}Y^n,
  \end{split}
  \label{zeta equation}\\
%end{equation}
%begin{equation*}
    \mathbf{z}_\zeta = \big\lbrace z_{ij} \big\rbrace_{0\, 0}^{n\, i}\,, \quad \quad   z_{\sun} = -z_{00}.
    \notag
\end{gather}
Here $\mathbf{z}_\zeta$ is the vector of $M= (n+2)(n+1)/2$ parameters of the model.
Function~\eqref{zeta equation} represents the average surface of the Galactic disk,
defined by the spatial distribution of objects (in the general case of matter) of the
selected Galactic subsystem.
The distance $\rho$ from the object to the surface $\zeta_n(R,Z)$ along the normal to this
surface will be considered as the value of the deviation of the object from the model
average surface of the disk. 

To obtain an estimate of the vector~$\mathbf{z}_\zeta$, we
generally use the maximum likelihood method. In this paper, to simplify, we assume that
$\rho$ as a random variable is distributed according to a normal law with zero mean, that
is, that the probability density of $\rho$ has the form
\begin{equation} \label{probdensrho}
  f(\rho) = \frac{1}{\sigma_{\rho} \sqrt{2\pi}}\ e^{-\frac{\rho^2}{2 \sigma_{\rho}^2}},
\end{equation}
where $\sigma_{\rho}$ is the standard deviation. We assume that the value of
$\sigma_{\rho}$ is the same for all objects in the sample. However, in future work we
propose to investigate and take into account the dependence of $\sigma_{\rho}$ and/or
other dispersion parameters on the Galactocentric distance. With probability density
\eqref{probdensrho}, the likelihood function~$L$ and logarithmic likelihood function~$\cal
L$ are
\begin{gather}
    L = \prod\limits_{i = 1}^{N} \frac{1}{\sigma_{\rho}\sqrt{2\pi}}\,
	e^{-\frac{\rho_i^2}{2\sigma_{\rho}^2}}, \notag\\
    {\cal L} \equiv -\ln{L} 
	= \frac{N}{2}\ln{(2\pi)} + N\ln{\sigma_{\rho}}
	 +\frac{1}{2} \sum\limits_{i = 1}^N \frac{\rho_i^2}{\sigma_{\rho}^2}\,,
	\label{mle main}\\
    \rho_i=\rho_i(r_i,l_i,b_i;\mathbf{z}_\zeta),\notag
\end{gather}
where $N$ is the number of objects in the sample; $r_i$, $l_i$ and $b_i$ are the
heliocentric distance, galactic longitude and latitude of $i$-th object, correspondingly.
Being a dispersion parameter under parametrization~\eqref{probdensrho}, $\sigma_\rho$ is
included in the general vector of the problem
parameters~$\mathbf{z}=(\sigma_\rho,\mathbf{z}_\zeta)=(z_1,z_2,\:\ldots,\: z_K)$, where
$K=M+1$.  The minimum of the function~$\cal L$ gives estimates of the parameters,
including the value of the standard deviation~$\sigma_{\rho}$ of objects across the disk,
which represents the contributions of both the true (natural) vertical dispersion of
objects and the random uncertainty of distance estimates (the latter contribution is
negligible, see Section~\ref{sect:discussion}). Thus, under
assumption~\eqref{probdensrho}, the maximum likelihood method was reduced to the nonlinear
least squares method (Eq.~\eqref{mle main}). The resulting value of $\sigma_\rho$ was
multiplied by the coefficient $\sqrt{N/(N-M)}$ to obtain an unbiased estimate.  The
vector~$\mathbf{z}_{\text{err}}$ of mean parameter errors and the mean model prediction
error~$\sigma_{\zeta_n}(X, Y)$ were calculated based on the Hessian~$H({\cal L})$ with
elements $h_{ij}$ (\citealt{Hudson1964,Wall+2012}):
\begin{equation} \label{mle hesse matrix and z err}
  \begin{gathered}
 h_{ij} = \frac{\partial{\cal L}^2}{\partial {z}_i \partial {z}_j}\,, \enskip i = 1,\:\ldots,\:K,\enskip 
	j = 1,\:\ldots,\: K, \qquad    C = [H({\cal L})]^{-1}, \\
    \mathbf{z}_{\text{err}} = \left\lbrace \sqrt{c_{ii}} \right\rbrace_{i = 1}^K\,, \quad
%   \sigma_{\zeta_n}(X, Y) = \sigma_{\rho} \sqrt{\mathbf{a}_n^{\:T}(X, Y)\,C\,\mathbf{a}_n(X, Y)}.
    \sigma^2_{\zeta_n}(X, Y) = \{\grad[\zeta_n(X,Y;\mathbf{z}_\zeta)]\}^{\text{T}}\, C'\,
	\grad[\zeta_n(X,Y;\mathbf{z}_\zeta)],
  \end{gathered}
\end{equation}
where $c_{ii}$ are diagonal elements of the covariance matrix $C$, $C'$ is a submatrix of
$C$ that does not contain covariances involving $\sigma_\rho$; here, an estimate of the
vector~$\mathbf z$ obtained by minimizing~${\cal L}$ is substituted for all values.

After finding the parameters, an outlier exclusion algorithm described in
\citet{Nikiforov2012} was applied to the sample of objects under consideration.
This algorithm differs from the usual $3\sigma$ criterion in that it uses a 
variable exclusion limit, which increases as the number of objects increases.

In order to check how well the observed distribution of sample objects by deviations
$\rho$ agrees with the model probability density function~\eqref{probdensrho}, we use
Pearson's chi-square test.

A priori choice of the order $n$ of expansion~\eqref{zeta equation} would lead to
significant errors in all parameters, including $z_{00}$ (see Sect.~\ref{sect:Results}),
so the value of~$n$ was also optimized using a simple algorithm. Models of the order
$n=0$, 1, 2 and so on are built sequentially. For each model of order $n$, the number of
parameters~$z_{ij}$ of order $n$ ($i + j = n$) whose estimates differ from zero at the
significance level ${\ge}2\sigma$ ($\sigma_{z_{ij}} /|z_{ij}| \le 0.5$) is calculated.
Then we find a model of the highest order such that it has at least one
$2\sigma$-significant parameter of the same order as the model. If the total number of
significant parameters of the selected model is greater than the corresponding number for
any lower-order model, then the selected model of order $n$ is assumed to be optimal.
Otherwise, a model of order $n-1$ is considered as possibly optimal and is compared in the
same way with models of lower orders in terms of the number of significant parameters.
Then either the order~$n-1$ is assumed to be optimal, or a transition is made to the
order~$n-2$, and so on until some order $n\ge 0$ is accepted as optimal, $n_\text{o}$. The
importance of choosing the correct model order will be illustrated in
Section~\ref{sect:Results}.

\section{Data}
\label{sect:Data}

We use the catalog of classical Cepheids by \citet{Berdnikov+2000} in the version of
\citet{Mel'nik+2015}, which provides data for 674 Cepheids from the General Catalogue of Variable
Stars (e.g., \citealt{Samus+2017}). The catalog is an updated version of the catalog
of Cepheid parameters by~\citet{Berdnikov+2000}. Observational data were obtained with
0.4--1 m telescopes of the Maidanak Observatory (Republic of Uzbekistan), Cerro Tololo and
Las Campanas observatories (Chile), Cerro Armazones Observatory of the Catholic University
(Chile) and South African Astronomical Observatory (see \citealt{Dambis+2015} for
details). In particular, in order to reliably determine the distances, Cepheids discovered
during the CCD monitoring of the southern sky performed as a part of the All Sky Automated
Survey (ASAS) project (\citealt{Pojmanski2002}) were observed, therefore, a survey of
Cepheids across the entire sky was performed.  The peak of the distribution of Cepheids by
the mean apparent magnitudes in the $V$ band falls on the values of $\langle
V\rangle=12$--$13$~mag, the limiting magnitude of the catalog is $\langle
V\rangle=15$~mag. The distances were obtained based on  the period--luminosity relation in
the $K$ infrared band and interstellar-extinction law using the period~-- normal color $(B
- V)$  relation derived earlier (see \citealt{Dambis+2015}).

%on equatorial and galactic coordinates, line-of-sight velocities, proper motions,
%pulsation periods and apparent magnitudes of 

%The line-of-sight velocities were taken from several sources,
%the proper motions were adopted from the reduction of Hipparcos data
%by}~\citet{van+Leeuwen2007}.

The authors of the catalog did not specify the possible value of the distance modulus
error. However, the analysis of these data in \citet{Veselova+Nikiforov2020} showed
that the nominal random errors of distance estimates given in \citet{Mel'nik+2015} are
small (the mean error of distance moduli $\sigma_d<0.14^m$). Using this distance catalog
gives us the opportunity to apply a simpler method that does not take into account random
distance errors. In the future, we intend to use newer data covering more extensive areas
where taking into account the uncertainty of distances within a more complex method is
necessary. Recent versions of \citeauthor{Berdnikov+2000}'s catalog have been successfully
used to study the Galactic structure. Based on the catalog, \citet{Mel'nik+2015}
identified signs of ring formations in the Galaxy, \citet{Dambis+2015} and
\citet{Veselova+Nikiforov2020} performed spatial modeling to determine the parameters of
spiral arm segments.

 According to the original distance
scale calibration of the catalog the distance to the Large Magellanic Cloud (LMC) is
$d^*_\text{LMC} = 18.25 \pm 0.05\,\text{mag}$ (\citealt{Berdnikov+2000}). Modern LMC
calibration is $d_\text{LMC} = 18.49 \pm 0.09\,\text{mag}$ (\citealt{deGrijs+2014}), which
leads to a correction factor~$c$ for the distances of the catalog used:
\begin{equation} \label{dist scale}
  c=\frac{r + \Delta r}{r} = 10^{0.2\Delta d} = 1.117^{+0.053}_{-0.050},\qquad 
 \Delta d = d_\text{LMC} - d^*_\text{LMC}.
\end{equation}
We analyzed the original catalog estimates of distances, and then adjusted the main
results for the factor $c$.

\begin{figure}
  \centering
  \includegraphics[width=10cm, angle=0]{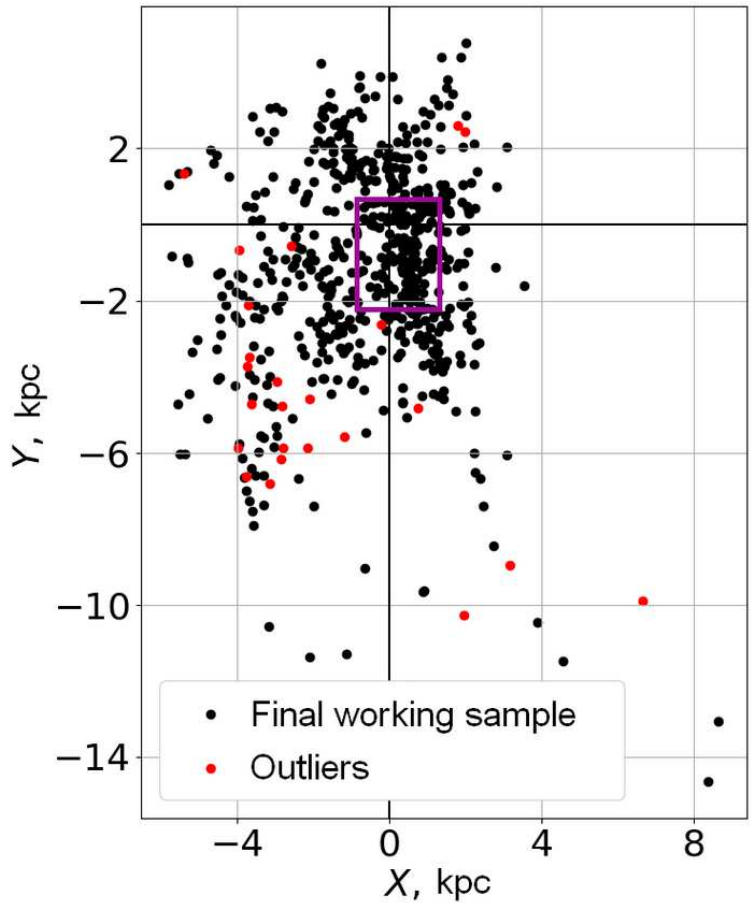}
  \caption{Objects of the final working sample and outliers projected onto the $XY$ plane.
   The purple rectangle marks the boundary of the local sample (see text).
   $X$-axis is directed towards the Galactic center, $Y$-axis is towards
   the rotation of the Galaxy. The Sun is placed at $X=0$~kpc, $Y=0$~kpc.} 
 \label{Cepheids_XY}
\end{figure}

Spatially isolated objects were manually excluded from the initial sample, i.e., those that
are nominally located clearly far from the main group of catalog objects. The remaining
objects, the totality of which we called the {\em working sample}, were used for
calculations. After applying the outlier elimination algorithm (\citealt{Nikiforov2012})
during the analysis of the working sample, we obtained the {\em final
working sample\/} containing 615 objects (Fig.~\ref{Cepheids_XY}).

\begin{figure}
  \centering
  \includegraphics[width=7.3cm, angle=0]{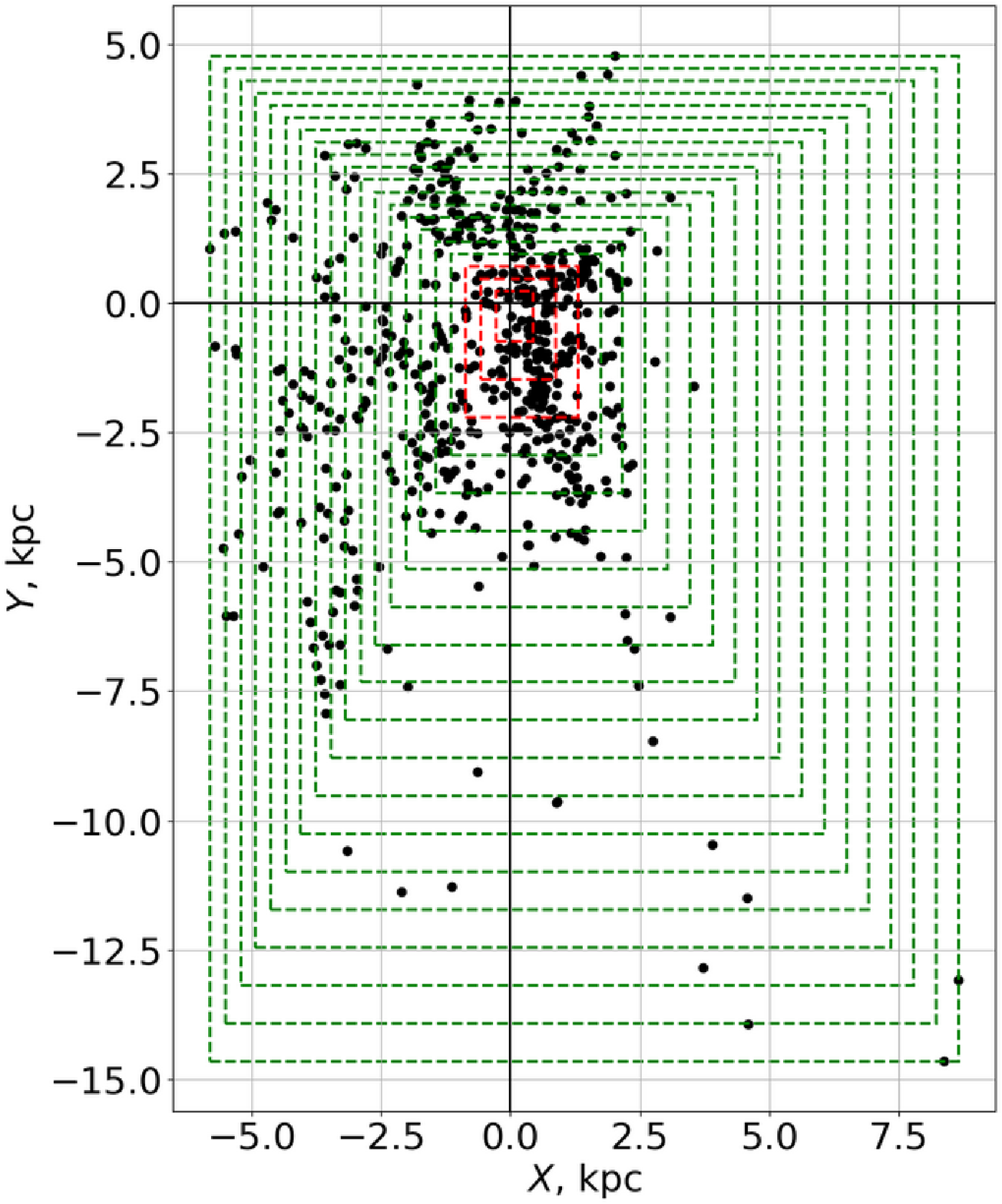}
  \includegraphics[width=7.3cm, angle=0]{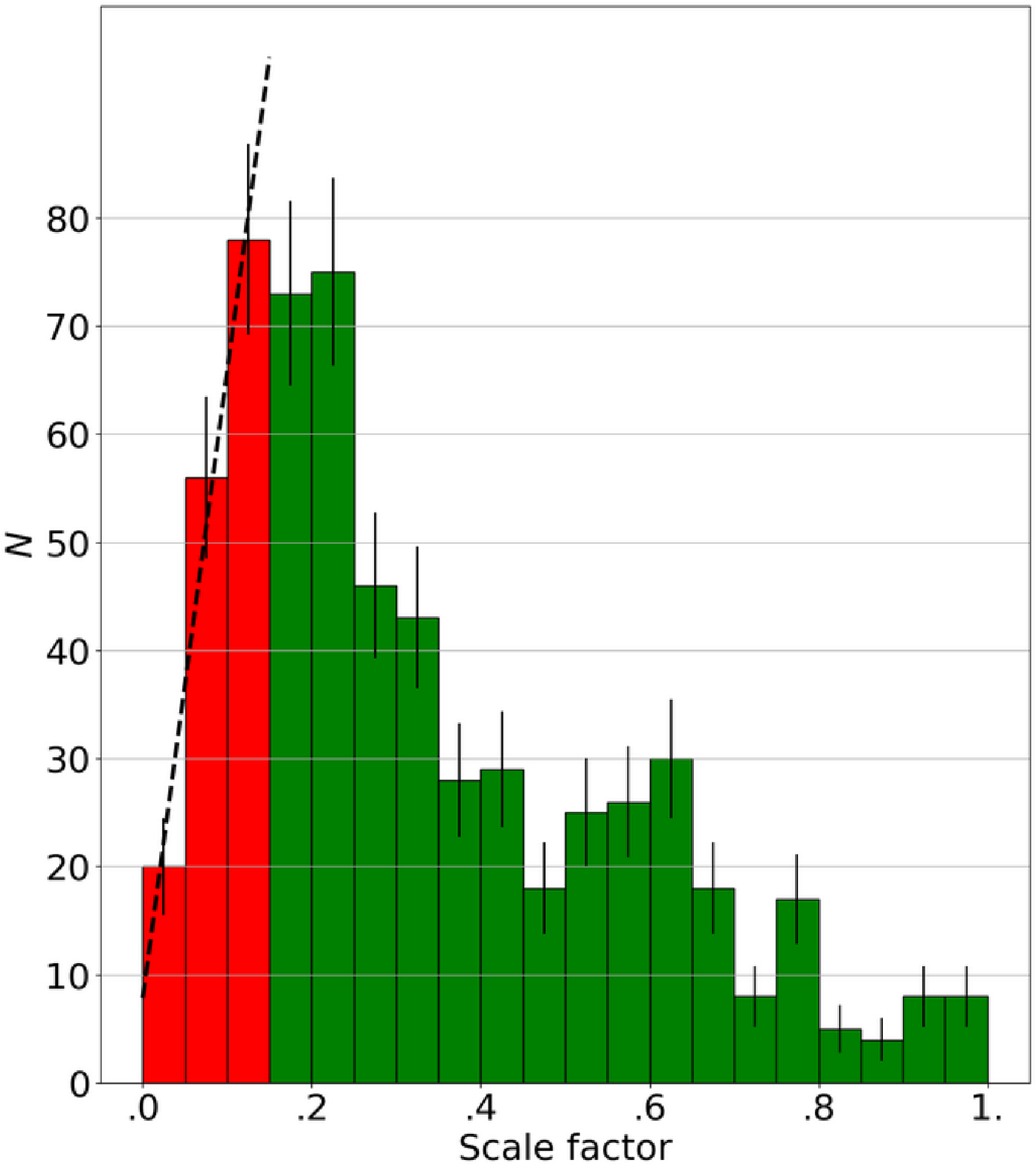}
 \caption{{\em Left\/}: Homothetic boundaries of rectangular frames used to construct the
 local sample (see text), superimposed on the distribution of objects of the working sample in
 projection on the $XY$ plane. {\em Right\/}: Histogram of the distribution of objects of the
 working sample in rectangular frames. The red color corresponds to the frames within which the
 sample (local sample) can be considered complete; the dashed
 straight line shows the linear growth area of the number of objects in the first three bins.
 }
  \label{Cepheids_rects+bars}
\end{figure}

We also identified a {\em local sample\/}, which is part of the working sample
representing the largest neighborhood of the Sun with the property of relative
completeness of the identification of objects of this type (classical Cepheids). Assuming
a homogeneous distribution of objects projected onto the $XY$-plane (the Galactic plane)
in a vicinity of the Sun, the number of objects projected onto a section of the $XY$-plane
should be proportional to the area of this section.  Usually, a set of concentric rings is
used to select a local, close to complete, subsample, but the working sample has an
asymmetry with respect to the origin of coordinates on the $XY$ plane. For this reason, it
was decided to use a set of rectangular frames with borders homothetic to the rectangular
border of the working sample: the lengths of each side of the frame vary according to a
given scale factor, and the ratio of distances from the origin to the edges of the frames
remain constant (Fig.~\ref{Cepheids_rects+bars}, left panel). In the case of a small frame
width, the area of the frame will be proportional to the outer perimeter, so for a
complete sample, the number of objects in the frame should increase linearly with the size
of the frame (scale factor). On the right panel of Fig.~\ref{Cepheids_rects+bars} the
linear growth  is observed within the first three bins (frames), which are marked in red;
objects within these frames were taken as a local sample. The boundaries of this sample
are shown in  Fig.~\ref{Cepheids_XY} (purple rectangle). The sample size is  154. After
excluding an outlier, the {\em final local sample\/} of 153 objects was obtained.

\section{Results}
\label{sect:Results}
\begin{figure}[t]
  \centering
  \includegraphics[width=10cm, angle=0]{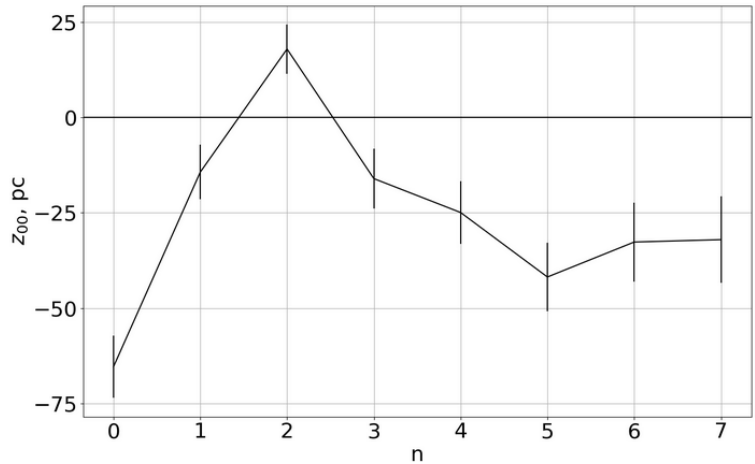}

  \caption{Dependence of $z_{00}$ (${=}-z_{\sun}$) on the model order $n$ for the final
  working sample. Vertical bars show the uncertainties of estimates.}%$

 \label{fig_z00}
\end{figure}

The influence of the accepted order of the model $\zeta_n$ on the results can be illustrated by
the example of the dependence of $z_{00}$ on $n$ for the final working sample
(Fig.~\ref{fig_z00}).  It can be seen that for small orders the estimates of $z_{00}$
strongly depend on $n$, while for larger orders the estimates vary insignificantly.

The optimal order of the model $\zeta_n$ for the final working sample turned out to be
$n_\text{o}= 4$. Estimates of model parameters are presented in
Table~\ref{optimal_results}, significant estimates are given in bold. For the local
and final local samples, $n_\text{o}$ is~$0$. Final results for $z_{\sun}$ and
$\sigma_{\rho}$ without and with the correction of distance scale (\ref{dist scale}) for
the final working and final local samples are listed in Table~\ref{final_results}. Comparison
with Figure~\ref{fig_z00} shows that the choice of an underestimated order
($n<n_\text{o}$) of the model can lead to an obviously incorrect estimate of~$z_{\sun}$
(here at $n=0$, 2).

\begin{table}
      \caption{Parameter estimates for the model $\zeta_n$ of the average surface of
      the Galactic disk of optimal order $n_\text{o}=4$ obtained for the final working
      sample of classical Cepheids. Estimates that differ from zero at a
      significance level of at least $2\sigma$ are highlighted in bold.
      The standard deviation ${\sigma_{\rho}}$ is given in
      kiloparsecs, while values of $z_{ij}$ are in units of $\text{kpc}^{1
      - i - j}$}
    \label{optimal_results}
      \begin{center}
      \begin{tabular}{ccc|ccc}
      \hline
      Parameter & Estimation & $\sigma_{z_i}/|z_i|$ & Parameter & Estimation & $\sigma_{z_i}/|z_i|$ \\
       \hline
      $\sigma_{\rho}$  &  \boldmath$0.1187 \pm 0.0033$  &  \boldmath$0.03$   &  $z_{21}$  &  \boldmath$0.00289 \pm 0.00094$  &  \boldmath$0.33$\\ 
%      \hline 
      $z_{00}$  &  \boldmath$-0.0243 \pm 0.0079$  &  \boldmath$0.33$ &  $z_{12}$  &  \boldmath$0.00288 \pm 0.00055$  &  \boldmath$0.19$\\ 
%      \hline 
      $z_{10}$  &  \boldmath$-0.0320 \pm 0.0061$  &  \boldmath$0.19$ &   $z_{03}$  &  \boldmath$0.00180 \pm 0.00028$  &  \boldmath$0.16$\\ 
%      \hline 
      $z_{01}$  &  \boldmath$-0.0131 \pm 0.0040$  &  \boldmath$0.31$ & $z_{40}$  &  $-0.00012 \pm 0.00012$  &  $1.00$\\ 
%      \hline 
      $z_{20}$  &  \boldmath$0.0055 \pm 0.0025$  &  \boldmath$0.45$ &  $z_{31}$  &  $0.00013 \pm 0.00014$  &  $1.08$\\
%       \hline 
      $z_{11}$  &  $0.0008 \pm 0.0018$  &  $2.25$ &  $z_{22}$  &  \boldmath$0.00022 \pm 0.00010$  &  \boldmath$0.45$\\
%       \hline 
      $z_{02}$  &  $-0.0007 \pm 0.0010$  &  $1.43$ &  $z_{13}$  &  \boldmath$0.000185 \pm 0.000048$  &  \boldmath$0.26$\\
%       \hline 
      $z_{30}$  &  \boldmath$0.00203 \pm 0.00087$  &  \boldmath$0.43$ &    $z_{04}$  &  \boldmath$0.000099 \pm 0.000020$  &  \boldmath$0.20$\\ 
      \hline 
      \end{tabular}
      \end{center}
    \end{table}

\begin{table}
  \caption{Final estimates of $z_{\sun}$ and $\sigma_{\rho}$ for the optimal
  models~$\zeta_{n_\text{o}}(X,Y)$ obtained for two samples of Cepheids. The estimates are given in
  the original catalog distance scale and in the scale adjusted for calibration
  $d_\text{LMC} = 18.49 \pm 0.09\,\text{mag}$ (see text)}%$
  \label{final_results}
  \begin{center}
    \begin{tabular}{l|cc}
      \hline
      Sample        & Original distance scale                    & Corrected distance scale\\
      \hline
      Final working sample& $z_{\sun} = 24.3 \pm 7.9\ \text{pc}$       & $z_{\sun} = 27.1 \pm 8.8\big|_{\text{stat.}}^{}\:^{+1.3}_{-1.2}\big|_{\text{cal.}}\: \text{pc}$       \\ %\cline{2-3}
      ($N=615$, $n_\text{o}=4$)& $\sigma_{\rho} = 118.7 \pm 3.3\ \text{pc}$ & $\sigma_{\rho} = 132.0 \pm 3.7\big|_{\text{stat}}^{}\:^{+6.3}_{-5.9}\big|_{\text{cal.}}\: \text{pc}$ \\ \hline
      Final local sample  & $z_{\sun} = 25.2 \pm 5.5\ \text{pc}$       & $z_{\sun} = 28.1 \pm 6.1\big|_{\text{stat.}}^{}\:\pm 1.3\big|_{\text{cal.}}\: \text{pc}$       \\ %\cline{2-3}
      ($N=153$, $n_\text{o}=0$)& $\sigma_{\rho} = 68.5 \pm 3.9\ \text{pc}$  & $\sigma_{\rho} = 76.5 \pm 4.4\big|_{\text{stat.}}^{}\:^{+3.6}_{-3.4}\big|_{\text{cal.}}\: \text{pc}$  \\ \hline
    \end{tabular}
  \end{center}
\end{table}

\begin{figure}[t!]
%  \centering
%  \vspace{-0.6cm}

%  \raisebox{1cm}{%
%  \hspace{.25cm}%
%%temp
\includegraphics[width=15cm, angle=0]{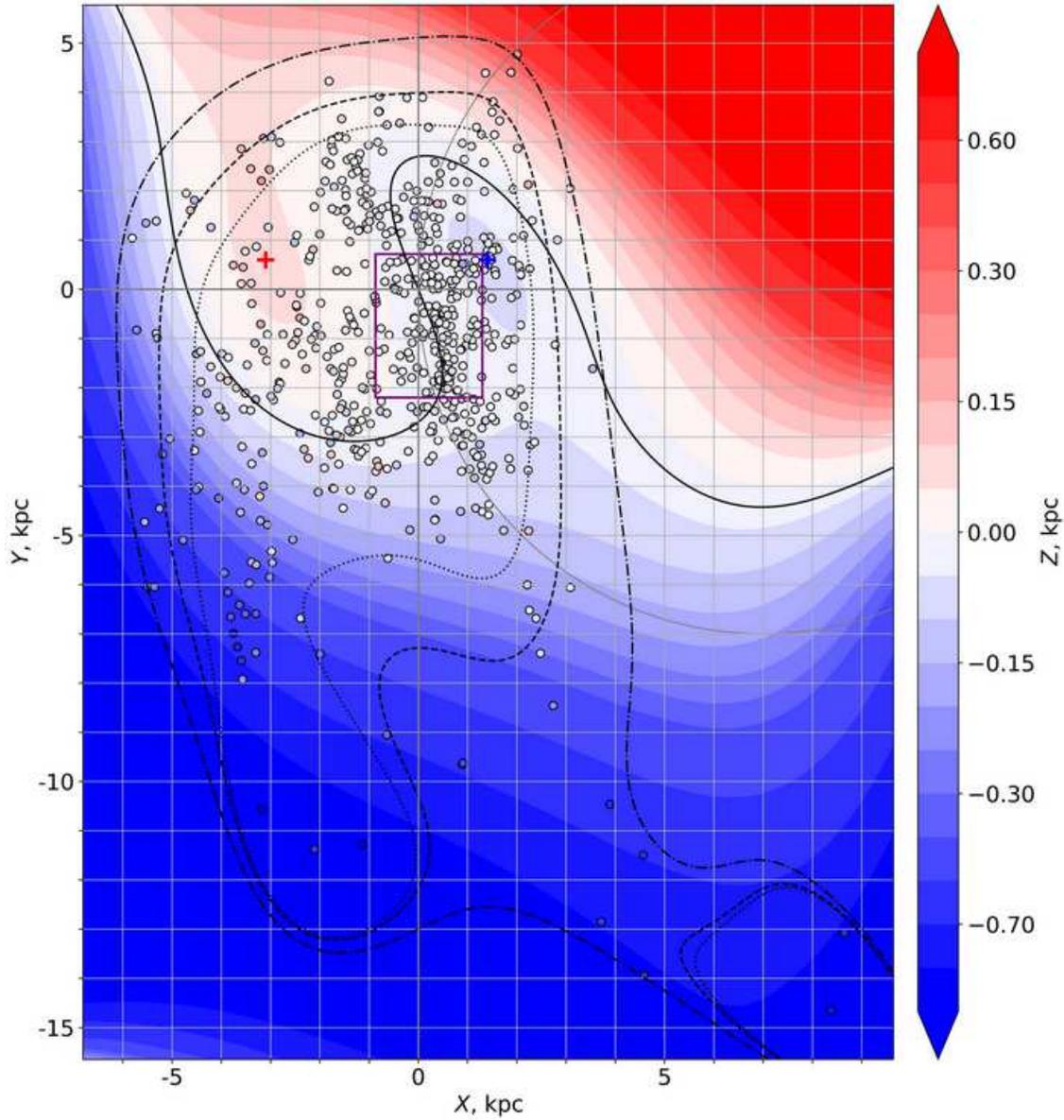}
% \includegraphics[width=15cm, angle=0]{ms2022-0209fig4.eps}
%\includegraphics[width=15cm, angle=0]{Cepheids_minM3_order4_trust_area_last.eps}
%  }%
%  \vspace{-2.5cm}%
  \caption{A map of the model $\zeta_4$ of the average disk surface of optimal
  order, constructed from the objects of the final working sample. The color shows the
  value of the function $\zeta_4(X,Y)$. The objects are shown as colored circles  with
  color representing the $Z$-coordinate. The black line represents the $z_{00}$-level, the
  red and blue crosses depict the local extrema (maximum and minimum, respectively), the
  purple rectangle shows the boundary of the local sample, the grey line denotes the solar
  circle $R=R_0=8$\,kpc. The dotted, dashed and dashed--dotted lines are the
  boundaries of the applicable areas of the model $\sigma_{\zeta_4}$, within which the
  mean error of the model is~$\sigma_{\zeta_4}(X,Y)\le
  \frac{1}{6} \sigma_\rho=20\,\text{pc}$, ${\le}\frac{1}{4} \sigma_\rho=40\,\text{pc}$ and 
  ${\le}\frac{1}{2} \sigma_\rho=59\,\text{pc}$, respectively.}
 \label{fig_map}
\end{figure}

The model surface $\zeta_4(X,Y)$ for the final working sample is shown in
Figure~\ref{fig_map}. Here the level 
\begin{equation}\label{curve-of-nodes}
 {\zeta_4}(X, Y) = z_{00}
\end{equation}
is represented by a black line.  This line is the intersection of the model of the average
surface of the disk with the nominal plane of the Galaxy $XY$. Line~\eqref{curve-of-nodes}
corresponds to the line-of-nodes in simple models. Figure~\ref{fig_map} shows that in
reality the line~\eqref{curve-of-nodes}, skirting the areas of local extrema, is quite
curved. By analogy with the line-of-nodes, the curve $\zeta_{n_\text{o}}(X,Y)=z_{00}$ can
be called a ``curve-of-nodes''. Of course, the constructed model~$\zeta_4(X,Y)$ is
real only for the area of the disk that is covered by the data used. Boundaries of areas
within which the mean error of the model is~$\sigma_{\zeta_4}(X,Y)\le
\frac{1}{6} \sigma_\rho=20\,\text{pc}$, ${\le}\frac{1}{4} \sigma_\rho=40\,\text{pc}$ and 
${\le}\frac{1}{2} \sigma_\rho=59\,\text{pc}$ shown in Figure~\ref{fig_map} by dotted,
dashed and dashed--dotted lines, respectively, give an idea of the applicability of the
model~$\zeta_4$ depending on the specified level of its uncertainty. Here
$\sigma_{\zeta_4}(X, Y)$ was calculated using the formula~\eqref{mle hesse matrix and z
err}, and $\sigma_\rho = 119$\,pc (see Table~\ref{optimal_results}).

\begin{figure}
  \centering
 \includegraphics[width=12cm, angle=0]{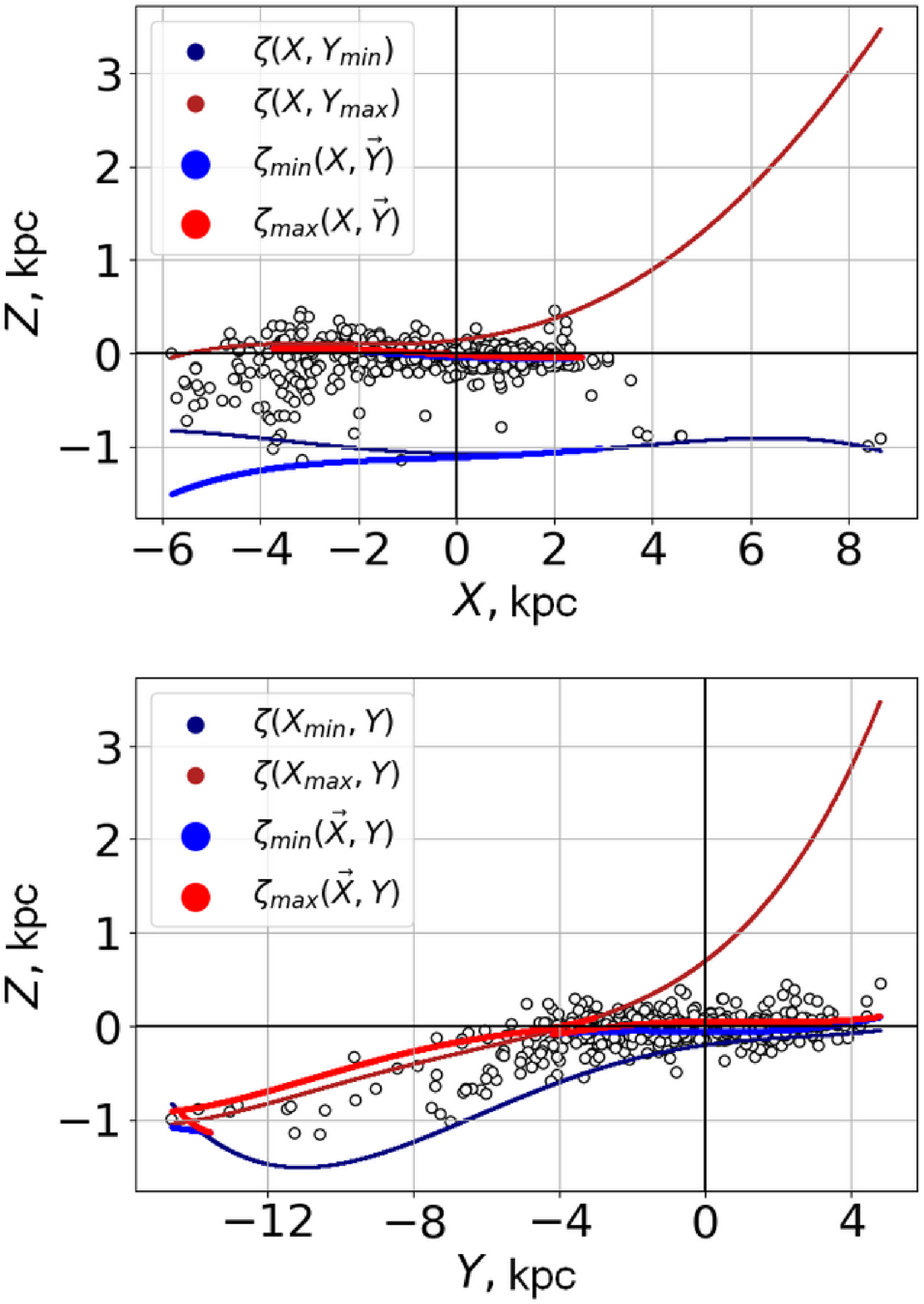}
  \caption{Extremum lines (thicker lines of lighter color) and boundary lines of the model
  surface $\zeta_4(X,Y)$ in comparison with Cepheids of the final working sample in the
  projection on the plane $XZ$ (top panel) and $YZ$ (bottom panel); see text. Local maxima
  and lines of the largest values of~$\zeta_4$ at the sample boundaries are shown in red, while
  local minima and lines of the smallest boundary values are shown in blue.}
 \label{fig_projection}
\end{figure}

An alternative representation of the resulting model is given in
Figure~\ref{fig_projection}, which shows the extremum and boundary lines of the model
surface $\zeta_4(X,Y)$ in projections on the planes $XZ$ and $YZ$ in comparison with the
Cepheids of the final working sample. The extremum lines (there may be several of them for
each projection) are dependencies of $Z$-coordinates of points of local extremes of the
surface~$\zeta_4(X,Y)$ with fixed $X$---$\zeta_\text{max}(X,\mathbf Y)$,
$\zeta_\text{min}(X,\mathbf Y)$ (Figure~\ref{fig_projection}, top panel) or with
fixed~$Y$---$\zeta_\text{max}(\mathbf X,Y)$, $\zeta_\text{min}(\mathbf X,Y)$ (bottom
panel). The boundary lines show the values of~$\zeta_4(X,Y)$ at the boundaries of the
final working sample: $\zeta_4(X,Y_\text{min})$ and $\zeta_4(X,Y_\text{max})$ (top panel),
and $\zeta_4(X_\text{min},Y)$ and $\zeta_4(X_\text{max},Y)$ (bottom panel), where
$X_\text{min}= -5.8$\,kpc, $X_\text{max}= 8.6$\,kpc, $Y_\text{min}= -14.6$\,kpc,
$Y_\text{max}= 4.8$\,kpc. Figure~\ref{fig_projection} affirms that the extremum lines
mainly fall on the areas covered by the data, and the boundary lines indicate edge
approximation effects in the area $X\gtrsim4$\,kpc, $Y\gtrsim0$\,kpc.

%The area of applicability of the model~$\zeta_4(X,Y)$ can also be judged by
%Figure~\ref{fig_projection},

The well-known general shape of the Galactic disk warp is clearly visible in
Figure~\ref{fig_map}---in the first and second quadrants, the model surface as a whole
rises above the $XY$ plane; in the third and fourth quadrants, the surface decreases below
this plane. In the area near $X\approx-3$ kpc, $Y\approx -6$ kpc, the decrease in the
average surface is observed for almost all objects.

\begin{figure}
   \centering
   \includegraphics[width=0.49\linewidth, angle=0]{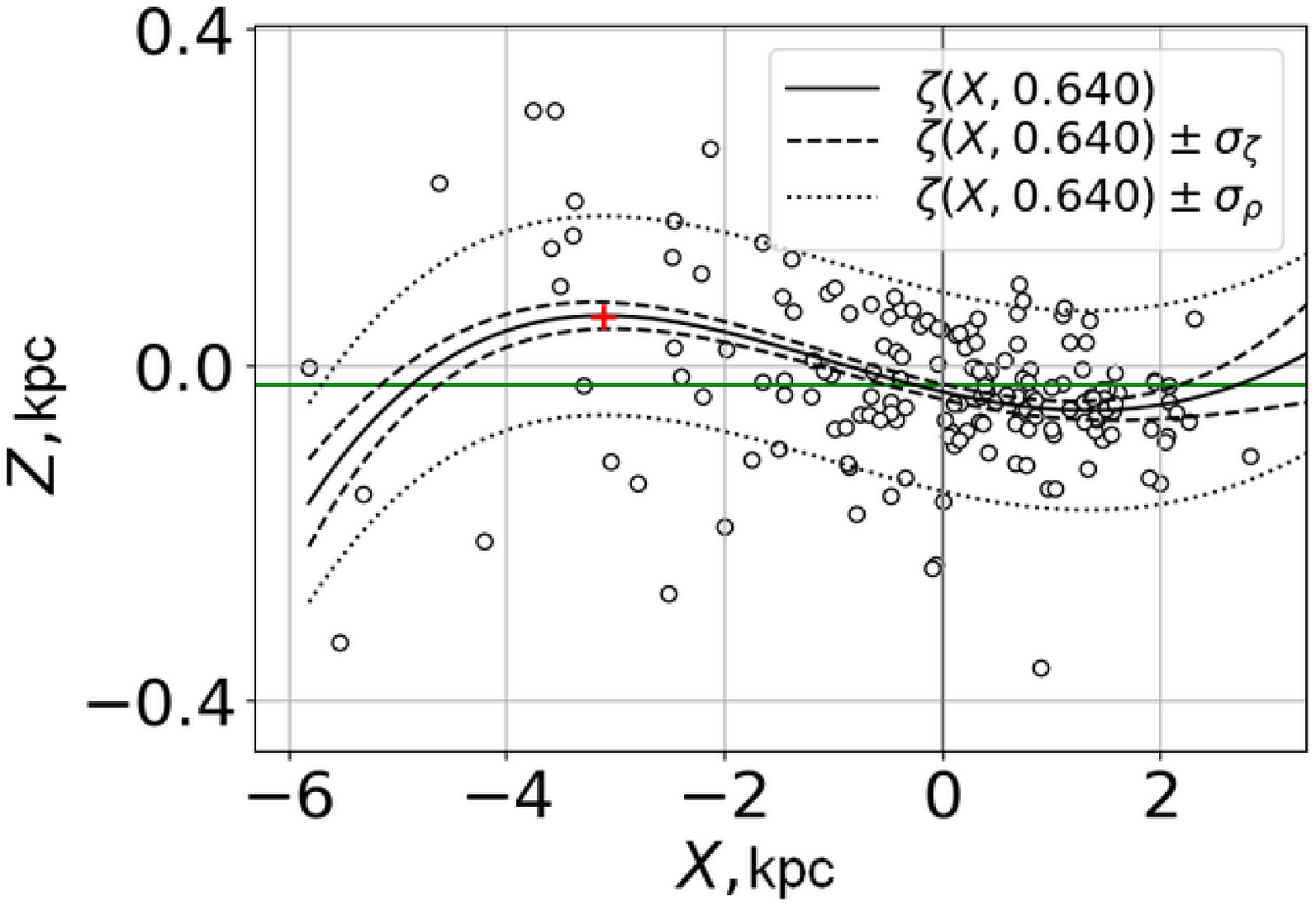}
   \includegraphics[width=0.49\linewidth, angle=0]{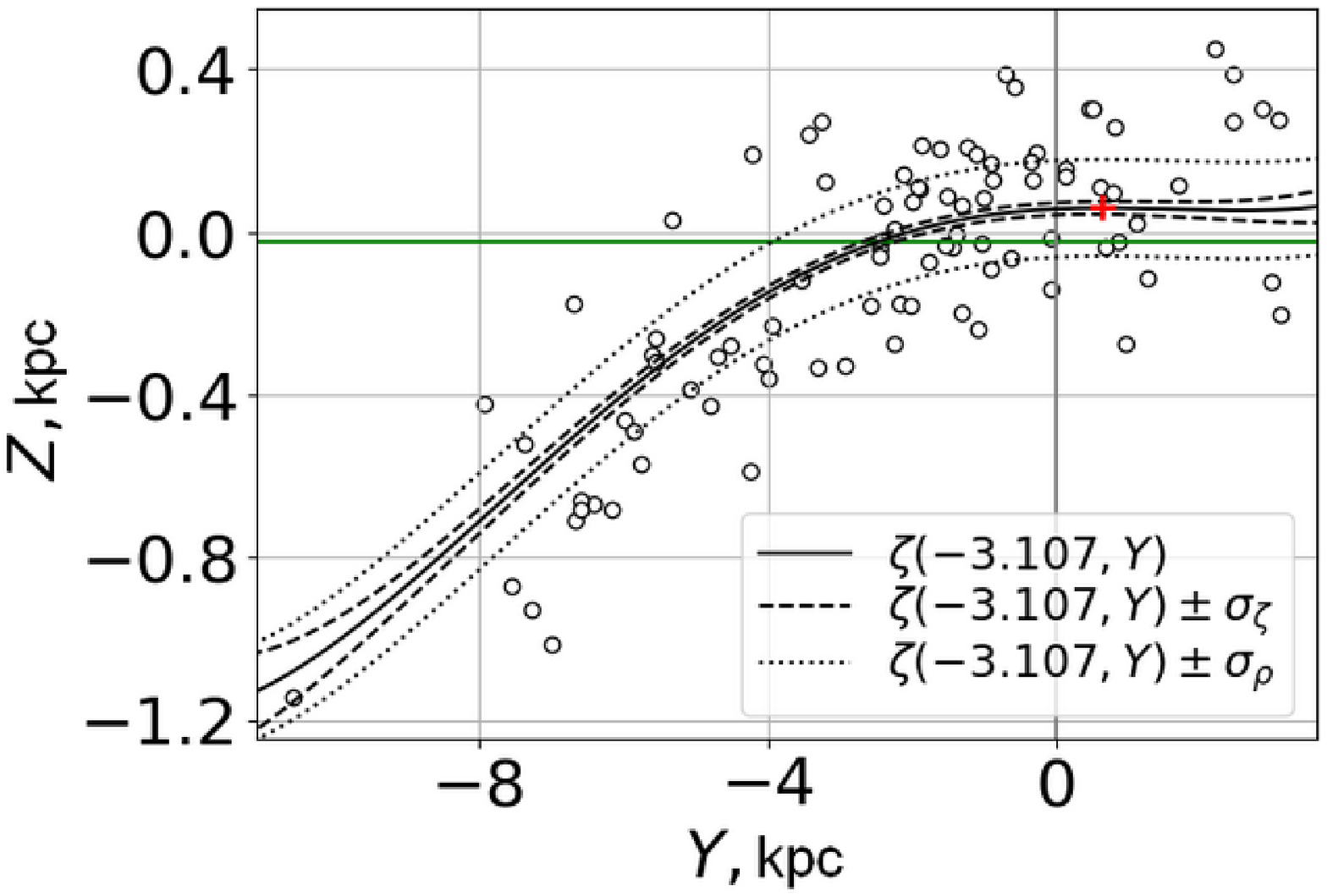}
   \raisebox{0.1cm}{%
   \hspace{0cm}%
   \includegraphics[width=0.49\linewidth]{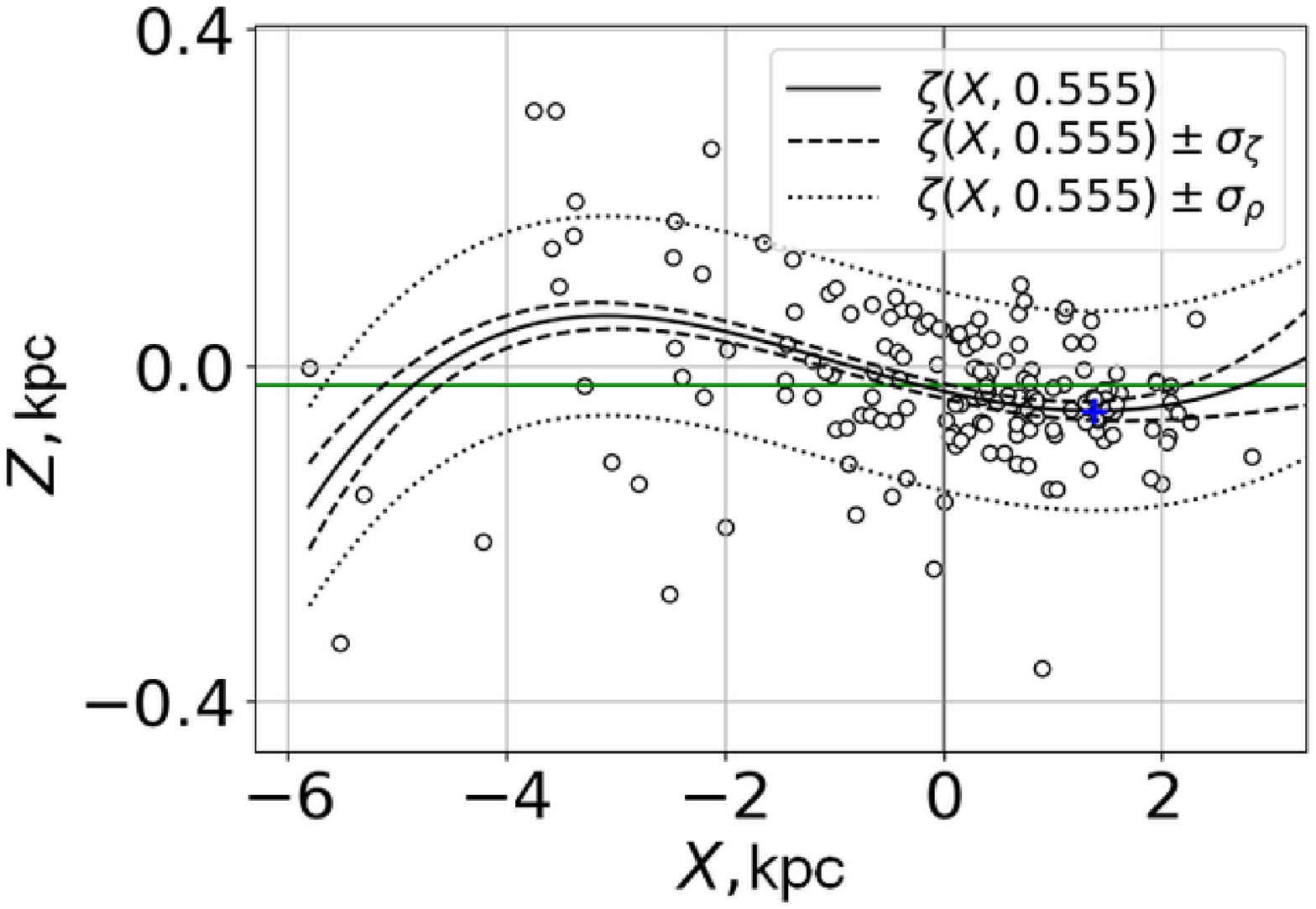}
   \includegraphics[width=0.49\linewidth]{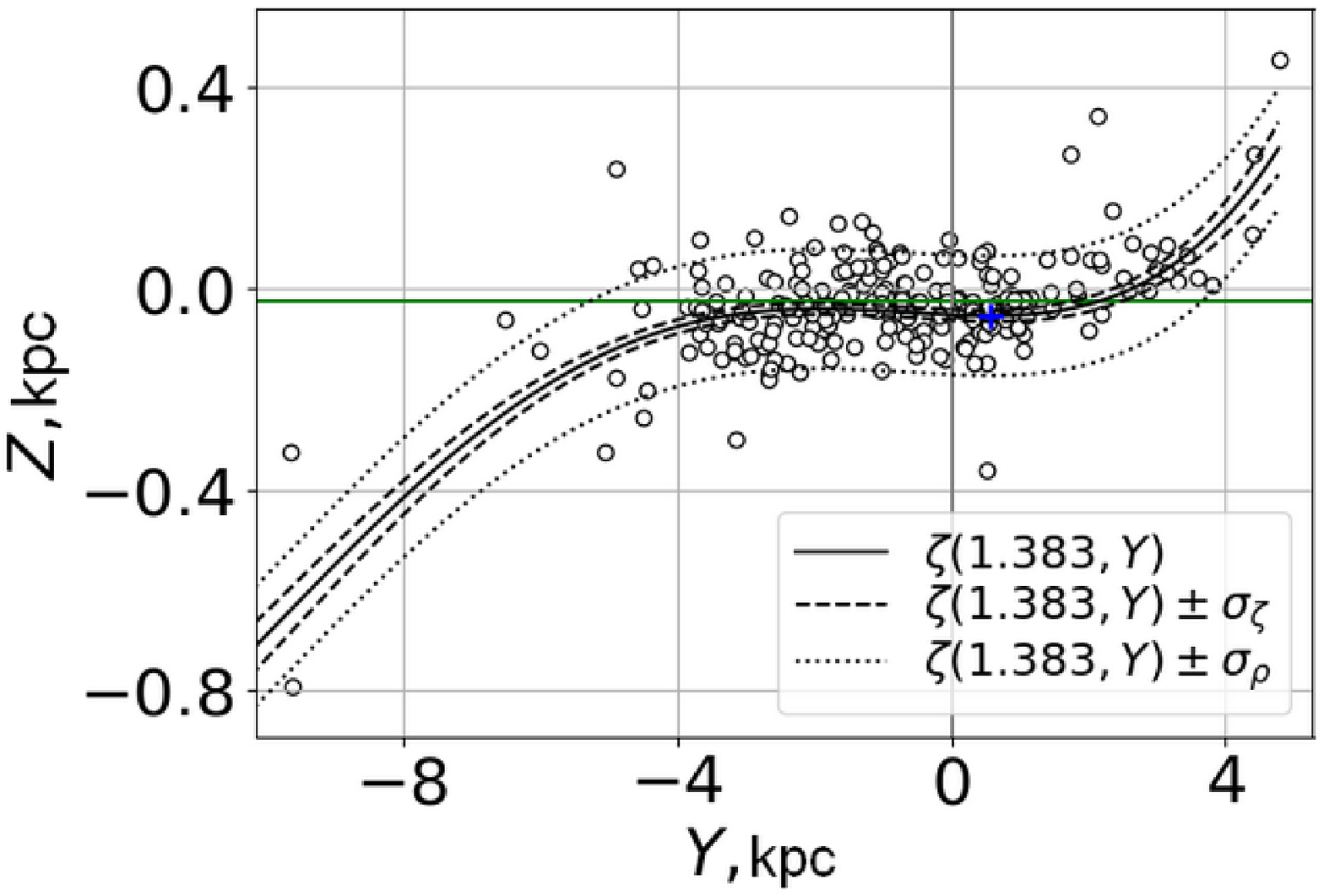}
   }%
   \caption{Sections of the model $\zeta_4(X,Y)$ of the middle surface of the disk passing through the
   local maximum (upper panels) and minimum (lower panels). Each panel shows the
   confidence area for the model values $\zeta_4(X,Y)$ ($\pm1\sigma_\zeta$) and the area
   of object deviations from the model $\pm 1\sigma_\rho$, as well as projections of the
   position of objects located in the $\pm 1\,\text{kpc}$ band from the cut line. The green
   line indicates the position of the plane $Z=z_{00}$.}
   \label{Cepheids_cuts}
\end{figure}

However, in addition, we found two extrema in the first and second quadrants that do not
fit into simple warp models.  The sections of the model  with planes parallel to
the $XZ$ and $YZ$ planes and passing through the points of extrema are shown in
Figure~\ref{Cepheids_cuts}.
The significance~$S$ of the extrema was estimated by the
formula
\begin{equation} \label{signif def}
S_{\zeta_4}(X, Y) = \frac{\zeta_4(X, Y) - z_{00}}{\sigma_{\zeta_4}(X, Y)}\,.
\end{equation}
Table~\ref{extrema} shows the parameters of local extrema. The given values of $S$ show
that the local minimum and maximum are significant at the level of at least $2\sigma$ and
$5\sigma$, respectively.
%The significance of the extrema of the average Galactic disk surface is shown
%for the first time in this paper.

\begin{table}
  \caption{Parameters of local extrema of the optimal model $\zeta_4(X,Y)$ constructed for
  the final working sample: The Cartesian heliocentric coordinates~$X,Y$ of the extremum,
  the value of $\zeta_4(X,Y)$ at the extremum point and  the significance level~$S_{\zeta_4}(X,
  Y)$ of the extremum (the number of $\sigma$)} \label{extrema}
  \begin{center}
    \begin{tabular}{c|cccc}
      \hline
      Extremum & $X$ (kpc) & $Y$ (kpc) & $\zeta_4(X,Y)$ (pc) & $S_{\zeta_4}(X, Y)$ \\
      \hline
      Local maximum & $-3.1$ & $0.6$ & $58.4 \pm 15.0$ & $5.43$ \\
      Local minimum & $1.4$ & $0.6$ & $-55.0 \pm 11.0$ & $-2.81$ \\
      \hline
    \end{tabular}
  \end{center}
\end{table}

In Figure~\ref{Cepheids_cuts} the slope of the model surface $\zeta_4(X,Y)$ to the nominal
plane of the Galaxy $XY$ is visible. Having calculated the distance~$\rho_{\sun}$ from the
Sun to the surface $\zeta_4(X,Y)$ along the normal to the latter, we obtained an estimate
of the angle of inclination of the local average surface of the Galactic disk to the plane
$XY$ of the galactic coordinate system
\begin{equation}\label{gamma angle}
 \gamma = \arccos\left(\frac{\rho_{\sun}}{z_{\sun}}\right) = 1\fdg79^{+ 0\fdg34}_{- 0\fdg33}.
\end{equation}
The $1\sigma$-uncertainty of the angle $\gamma$ indicated here was found by the Monte Carlo method
based on the results of processing 50 mock data catalogs.

\begin{figure}
  \centering
  \includegraphics[width=6.5cm, angle=0]{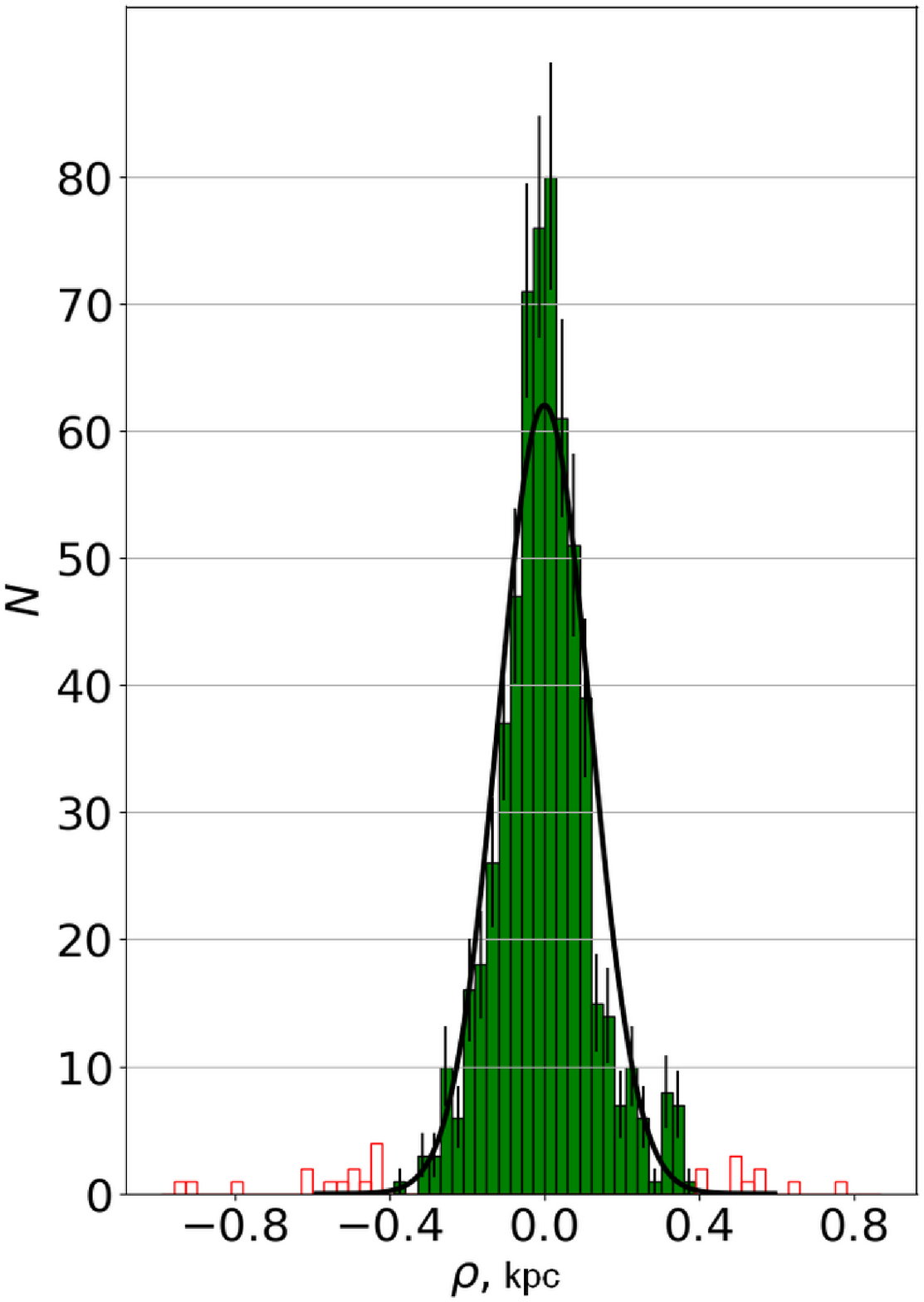}
  \raisebox{-0.29cm}{%
 \hspace{0.2cm}%
  \includegraphics[width=7.13cm, angle=0]{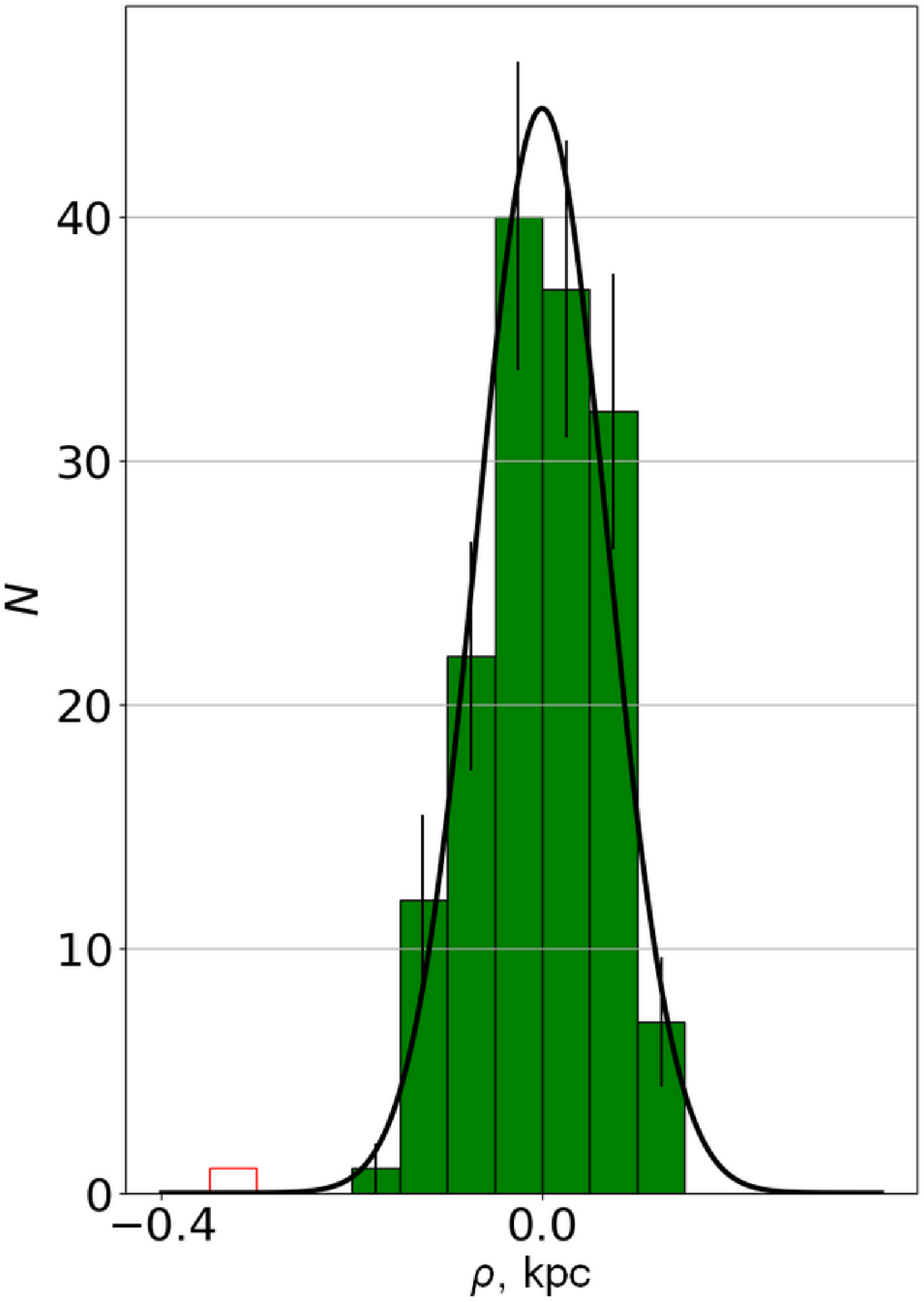}
}%

  \caption{Observed (green columns) and model (black line) distributions of object deviations $\rho$
  along the normal to the model average surface for the working sample (left panel,
  for $n_\text{o}=4$ and the model parameters given in Table~\ref{optimal_results})
  and local sample (right panel, for $n_\text{o}=1$ and the parameters given in
  Table~\ref{final_results}). Outliers are shown in red.}%$

 \label{fig_distrib}
\end{figure}

According to Pearson's chi-square test, the probability density function
$f(\rho)$~\eqref{probdensrho} does not fit data well (see Fig.~\ref{fig_distrib}, left
panel)---the probability of accepting the null hypothesis that the observed distribution
has a probability density of the form~\eqref{probdensrho} is less than $1\%$. This means
the other functions~$f(\rho)$ or combinations of them should be considered in the future.
However, for the final local sample the chi-square test gives a probability of
acceptance of the hypothesis~\eqref{probdensrho} of about $30\%$ (see
Fig.~\ref{fig_distrib}, right panel), i.e., the Gaussian distribution as a model for
$f(\rho)$ should not be excluded from consideration.

Since in this paper we used a catalog (\citealt{Mel'nik+2015}) based on observations in
infrared bands ($I_\text{C}$ and $K$; see \citealt{Dambis+2015}), this allows us to expect
a low selection effect due to the absorption of light by dust in the Galactic disk. In
particular, there should be no significant differential selection in the {\em vertical
direction}, i.e., statistical sample bias to the north of the disk. Indeed, due to the
position of the Sun above the average surface of the disk, a ray of light from an object
located south of the average surface of the disk passes through a larger thickness of the
disk and experiences greater light absorption. Indeed, due to the position of the Sun
above the average surface of the disk, a ray of light from an object located south of the
average surface of the disk passes through a larger thickness of the disk and experiences
greater light absorption compared to a northern object at the same distance from the
average surface. Therefore, in principle, one would expect a sharper truncation of the
observed distribution of deviations $\rho$ from the model from the negative $\rho$ side
compared to the positive ones, i.e., greater detection of northern objects compared to
southern ones. Asymmetry of this type in the distribution of deviations~$\rho$ does not
really manifest itself in a noticeable way (fig.~\ref{fig_distrib}). In reality, for the
working sample, instead of a sharper truncation on negative~$\rho$, rather on the
contrary, there is some deficit on~$\rho\sim+0.2$\,kpc (fig.~\ref{fig_distrib}, left
panel). The standard deviation calculated for objects with $\rho>0$\,pc for any sample
does not significantly exceed the standard deviation found for objects with $\rho<0$\,kpc
(see Table~\ref{sigma+-}). These results support of the insignificance of north--south
selection.%$

\begin{table}
  \caption{%
   Standard deviations of Cepheids across the disk, calculated
   for objects of the considered samples with only positive deviations~$\rho$ from the model,
   $\sigma_\rho^+$, and with only negative ones, $\sigma_\rho^-$%$
   }%
  \label{sigma+-}
  \begin{center}
    \begin{tabular}{l|cc}
      \hline
      Sample        & $\sigma_\rho^+$ (kpc) & $\sigma_\rho^-$ (kpc)\\
      \hline
      Working sample      & $0.1621 \pm 0.0045$ & $0.1641 \pm 0.0046$\\
      Final working sample& $0.1277 \pm 0.0053$ & $0.1162 \pm 0.0048$\\
      Local sample        & $0.0765 \pm 0.0043$ & $0.0894 \pm 0.0046$\\
      Final local sample  & $0.0757 \pm 0.0043$ & $0.0792 \pm 0.0043$\\
      \hline
    \end{tabular}
  \end{center}
\end{table}

Note that the increase in the selection effect with distance from the Sun in the sense of
incomplete sampling (Fig.~\ref{Cepheids_rects+bars}, right panel) does not in itself lead to bias,
i.e., to systematic errors in the position of the {\em average\/} (model) surface of the disk
and in estimation of the dispersion of objects relative to this surface. Classical Cepheids
belong to population~I and, being quite young objects---${\sim}10^7$--$10^8$ years (see, e.g.,
\citealt{Veselova+Nikiforov2020})--- represent a thin disk of the Galaxy. Moreover, the
vertical deviation of the subsystem of classical Cepheids $\sigma_{\rho}{\sim}130$\,pc (this
work) is significantly smaller than the average vertical scale $h_z=300\pm50$\,pc of the thin
disk as a whole (\citealt{Bland-Hawthorn+2016}). This makes classical Cepheids a good
tracer of the thin disk of the Galaxy: even in areas where there are few of them, they still
represent the position of the disk with a relatively small spread.   In such areas the
mathematical expectation of the average $Z$-coordinate of the Cepheids, $\overline Z$, remains
equal to the true $Z$-coordinate of the average surface of the disk, only the mean error of
$\overline Z$ estimate increases, i.e., in the case of our method, the mean error of the model
$\sigma_{\zeta_4}$ increases (isolines of $\sigma_{\zeta_4}(X,Y)$ are shown on
Figure~\ref{fig_map}).

On the other hand, Cepheids, concentrating towards spiral arms (see, e.g.,
\citealt{Veselova+Nikiforov2020}), like other tracers of the spiral structure of the Galaxy,
often have not a uniform, but a ``patchy'' distribution along the spiral arms (for example,
\citealt[fig.~1]{Efremov2011}; \citealt[fig.~13]{Nikiforov+Veselova2018a};
\citealt[fig.~2]{Reid+2019}; \citealt[fig.~5]{Veselova+Nikiforov2020}). This, as well as the
growth of incomplete detection of objects and a decrease in the density of the disk to the
periphery, leads to the fact that at large distances from the Sun, gaps appear in the
distribution of Cepheids, for example at $(X,Y)\sim(0, -7)$\,kpc (Figure~\ref{fig_map}). Of
course, in the areas of such gaps, the constructed model should be treated only as a smooth
interpolation of the average trend between the areas represented by the data. However, when
imposing a stricter restriction on the mean error of the model, e.g.,
$\sigma_{\zeta_4}(X,Y)\le\frac{1}{6}\sigma_\rho=20\,\text{pc}$, the internal area of
applicability of the model does not include most of these lacunae (see Figure~\ref{fig_map}).
The relatively high accuracy of the model for the area in the lower right corner of
Figure~\ref{fig_map} is, of course, only formal, due to the fact that any sufficiently flexible
model must pass through a few Cepheids in this area, located at some distance from the bulk of
the sample objects. On the other hand, these and other Cepheids at large negative $Y$ are in
the general trend of a well-known decrease in the average surface of the disk in this area. So,
in the region ~$Y< -10$\,kpc, all Cepheids of the working sample are in the range~$-1.2\le Z
\le-0.8$\,kpc (Figure~\ref{fig_projection}, bottom panel), which agrees well with the disk
level in this area according to other data---e.g., $-2\le Z\le 1$\,kpc
(\citealt{Skowron+2019b}, from Cepheids), $\overline Z= -1.22$~kpc
(\citealt{Romero-Gomez+2019}, from red giant branch (RGB) stars), $-1\le Z\le -0.5$\,kpc
(\citealt{Lemasle+2022}, from Cepheids).  The vertical dispersion of objects at large
negative~$Y$ is consistent with that for the rest of the working sample
(Figure~\ref{fig_projection}, bottom panel).  At the same time, the use of the entire
working sample, despite the gaps, does not create any fictitious ripples in the model of
the average surface of the disk in the area~$Y< -6$\,kpc (Figure~\ref{fig_map},
\ref{Cepheids_cuts}). Thus, there was no reason to discard Cepheids at large negative $Y$.
In addition, it was important to keep in the sample objects representing the decline of
the disk surface in the III and IV quadrants in order to test the capabilities of the
general model~\eqref{zeta equation} within the framework of the proposed method to
describe this well-known feature together with other possible details of the disk surface
(as it turned out, local extremes). Note that the value of the decrease in the disk level
of $\Delta Z{\sim}1$\,kpc relative to the plane $Z=z_{00}$ is much larger than the
standard deviation for Cepheids ($\sigma_{\rho} \sim 130$\,pc), i.e., the downward trend
is detected confidently, despite the incompleteness of the sample in this area.

\begin{figure}
  \centering
  \includegraphics[width=0.3\linewidth, angle=0]{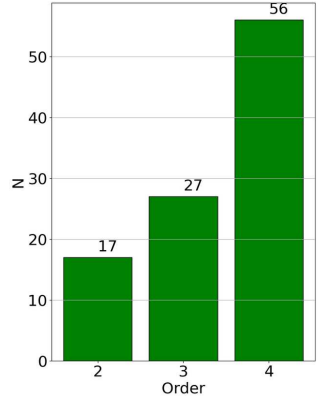}
  \caption{Distribution of optimal order values for 100 mock samples (see text).} 
 \label{MonteCarlo}
\end{figure}

We tested the algorithm used here for choosing the optimal order of the $\zeta_n$ model by
the Monte Carlo method. As a model of the disk surface, the constructed model $\zeta_4$
was adopted (Table~\ref{optimal_results}) and 100 mock catalogs were generated with object
deviations from $\zeta_4(X,Y)$ along the normal to this surface, distributed according to
the law~\eqref{probdensrho} with the value of $\sigma_\rho$ indicated in
Table~\ref{optimal_results}. The results are shown in Figure~\ref{MonteCarlo}. They show
that in most cases the order of the initial model is restored exactly, and the probability
that the order of the model will be underestimated is less than 1\%. These results suggest
that, acting according to this algorithm, it is possible to obtain a model that does not
fully reflect some details of the real disk structure, but it is unlikely to build an
excessively complex model with fictitious details.

\begin{table}
      \caption{Same as in Table~\ref{optimal_results}, but for the odd
      subsample of the final working sample (see text). $n_\text{o}=4$}
    \label{optimal_results_odds}
      \begin{center}
      \begin{tabular}{ccc|ccc}
      \hline
      Parameter & Estimation & $\sigma_{z_i}/|z_i|$ & Parameter & Estimation & $\sigma_{z_i}/|z_i|$ \\
       \hline
      $\sigma_{z}$  &  \boldmath$0.1218 \pm 0.0050$  &  \boldmath$0.04$& $z_{21}$  &  $0.0017 \pm 0.0013$  &  $0.76$                      \\ 
      $z_{00}$  &  \boldmath$-0.026 \pm 0.011$  &  \boldmath$0.42$     & $z_{12}$  &  \boldmath$0.00214 \pm 0.00080$  &  \boldmath$0.37$  \\ 
      $z_{10}$  &  \boldmath$-0.0321 \pm 0.0093$  &  \boldmath$0.29$   & $z_{03}$  &  \boldmath$0.00145 \pm 0.00040$  &  \boldmath$0.28$  \\ 
      $z_{01}$  &  $-0.0077 \pm 0.0061$  &  $0.79$                     & $z_{40}$  &  $0.00025 \pm 0.00022$  &  $0.88$                    \\ 
      $z_{20}$  &  $0.0074 \pm 0.0039$  &  $0.53$                      & $z_{31}$  &  $0.00026 \pm 0.00021$  &  $0.81$                    \\
      $z_{11}$  &  $-0.0002 \pm 0.0026$  &  $13.00$                    & $z_{22}$  &  $-0.00008 \pm 0.00018$  &  $2.25$                   \\
      $z_{02}$  &  $-0.0013 \pm 0.0014$  &  $1.08$                     & $z_{13}$  &  \boldmath$0.000133 \pm 0.000065$  &  \boldmath$0.49$\\
      $z_{30}$  &  \boldmath$0.0038 \pm 0.0015$  &  \boldmath$0.39$    & $z_{04}$  &  \boldmath$0.000092 \pm 0.000029$  &  \boldmath$0.32$\\ 
      \hline 
      \end{tabular}
      \end{center}
    \end{table}

\begin{figure}
   \centering
%%temp
   \includegraphics[width=0.49\linewidth, angle=0]{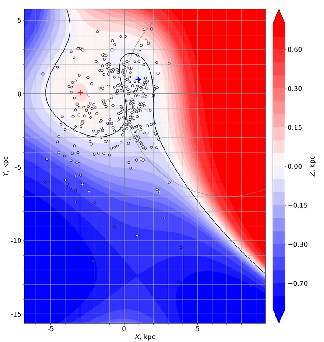}%
   \hspace{0.8em}%
   \includegraphics[width=0.49\linewidth, angle=0]{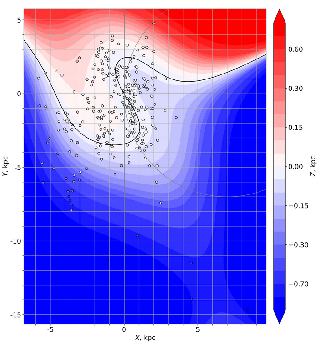}
%  \includegraphics[width=0.49\linewidth, angle=0]{ms2022-0209fig9a.eps}%
%  \hspace{0.8em}%
%  \includegraphics[width=0.49\linewidth, angle=0]{ms2022-0209fig9b.eps}
%  \includegraphics[width=0.49\linewidth, angle=0]{Cepheids_order4_odd.eps}%
%  \hspace{0.8em}%
%  \includegraphics[width=0.49\linewidth, angle=0]{Cepheids_order4_even.eps}
  \caption{Maps of the models $\zeta_4$ of the average disk surface of optimal
  order, constructed from the objects of independent odd (left panel) and even (right panel)
  subsamples of the final working sample (see text). The notation is the same as in
  Figure~\ref{fig_map}.}
   \label{fig_map:odd+even}
\end{figure}

To check the stability of the results, we divided the final working sample (hereinafter,
for short, ``full sample'') into two parts: one part included objects with odd numbers in
the sample list (we will call it the {\em odd subsample}), and the other objects with even
numbers ({\em even subsample}). This separation is actually random. On the other hand, it
keeps the relative population of data of different longitude intervals approximately the
same for both subsamples and for the full sample, since in the catalog
\citet{Mel'nik+2015} objects are ordered by their names, i.e., mainly by the names of
constellations. The latter is important if we want to check the reproducibility of the
detected details on the relief of the disk surface. The calculations were repeated for
each of the two independent subsamples. The results are presented in
Tables~\ref{optimal_results_odds}--\ref{extrema_even} and in
Figure~\ref{fig_map:odd+even}. The optimal orders of the model~$\zeta_n$ for both
subsamples turned out to be the same and equal to~$n_\text{o}$ for the full sample:
$n_\text{o}=4$. The parameters of the model~$\zeta_4$ for even and odd subsamples and for
the full sample within the error limits are consistent with each other
(cf.~Tables~\ref{optimal_results},
\ref{optimal_results_odds}, \ref{optimal_results_evens}). At the same time, all
significant parameters obtained from subsamples are also significant for the full sample.
The characteristics of the local extremes for the odd subsample turned out to be similar to
the characteristics for the full sample (cf.~Tables~\ref{extrema}, \ref{extrema_even}).
For the even subsample, the model~$\zeta_4(X,Y)$ does not formally have local extremes, but it
has areas of depression and elevation (Figure~\ref{fig_map:odd+even}, right panel), in
position and amplitude close to those for models obtained from the full sample and odd subsample
(Figure~\ref{fig_map}; Figure~\ref{fig_map:odd+even}, left panel). The curves-of-nodes 
also turned out to be similar for all three samples in the area of applicability of the
model (Figures~\ref{fig_map},
\ref{fig_map:odd+even}). Thus, the topology of the resulting model as a whole is preserved
even when the sample is divided. At the same time, the drop in the significance of the
details of the model surface~$\zeta_4(X,Y)$ for subsamples shows that dividing the sample
into a larger number of parts is hardly meaningful.

\begin{table}
      \caption{Same as in Table~\ref{optimal_results}, but for the even
      subsample of the final working sample (see text). $n_\text{o}=4$}
    \label{optimal_results_evens}
      \begin{center}
      \begin{tabular}{ccc|ccc}
      \hline
      Parameter & Estimation & $\sigma_{z_i}/|z_i|$ & Parameter & Estimation & $\sigma_{z_i}/|z_i|$ \\
       \hline
      $\sigma_{z}$  &  \boldmath$0.1227 \pm 0.0050$  &  \boldmath$0.04$& $z_{21}$  &  \boldmath$0.0037 \pm 0.0015$  &  \boldmath$0.41$    \\ 
      $z_{00}$  &  \boldmath$-0.028 \pm 0.011$  &  \boldmath$0.39$     & $z_{12}$  &  \boldmath$0.00249 \pm 0.00086$  &  \boldmath$0.35$  \\ 
      $z_{10}$  &  \boldmath$-0.0279 \pm 0.0097$  &  \boldmath$0.35$   & $z_{03}$  &  \boldmath$0.00210 \pm 0.00044$  &  \boldmath$0.21$  \\ 
      $z_{01}$  &  \boldmath$-0.0167 \pm 0.0060$  &  \boldmath$0.36$   & $z_{40}$  &  $-0.00030 \pm 0.00020$  &  $0.67$                   \\ 
      $z_{20}$  &  $0.0046 \pm 0.0036$  &  $0.78$                      & $z_{31}$  &  $0.00010 \pm 0.00020$  &  $2.00$                    \\
      $z_{11}$  &  $0.0011 \pm 0.0028$  &  $2.55$                      & $z_{22}$  &  \boldmath$0.00035 \pm 0.00016$  &  \boldmath$0.46$  \\
      $z_{02}$  &  $0.0011 \pm 0.0016$  &  $1.45$                      & $z_{13}$  &  $0.000125 \pm 0.000088$  &  $0.70$                  \\
      $z_{30}$  &  $0.0008 \pm 0.0014$  &  $1.75$                      & $z_{04}$  &  \boldmath$0.000097 \pm 0.000033$  &  \boldmath$0.34$\\ 
      \hline 
      \end{tabular}
      \end{center}
    \end{table}

\begin{table}
  \caption{Same as in Table~\ref{extrema}, but for the odd
      subsample of the final working sample (see text)}
  \label{extrema_even}
  \begin{center}
    \begin{tabular}{c|cccc}
      \hline
      Extremum & $X$ (kpc) & $Y$ (kpc) & $\zeta_4(X,Y)$ (pc) & $S_{\zeta_4}(X, Y)$ \\
      \hline
      Local maximum & $-3.0$ & $0.1$ & $53.0 \pm 20$ & $4.01$ \\
      Local minimum & $1.0$ & $1.0$ & $-50.4 \pm 15$ & $-1.59$ \\
      \hline
    \end{tabular}
  \end{center}
\end{table}

\section{Discussion}
\label{sect:discussion}

Consistency of local solar offset estimate $z_{\sun} = 28.1 \pm
6.1\big|_{\text{stat}}^{}\:^{+1.3}_{-1.3}\big|_{\text{cal}}\: \text{pc}$ and global
estimate $z_{\sun} = 27.1 \pm
8.8\big|_{\text{stat}}^{}\:^{+1.3}_{-1.2}\big|_{\text{cal}}\: \text{pc}$
(Table~\ref{final_results}) also implies that the proposed algorithm is valid and the
model order is correct---when optimizing the order of the model, the value of the
solar offset does not depend systematically on whether a large or small neighborhood of the Sun is
considered. Our estimates also agree with the best estimate of the solar offset $z_{\sun}
= 25 \pm 5\,\text{pc}$
\citep{Bland-Hawthorn+2016}
and local estimates specifically for classical Cepheids (see below).

However, the estimate of $\sigma_{\rho} = 68.5 \pm 3.9$\,pc obtained for the
final local sample is inconsistent with the estimate of $\sigma_{\rho} = 118.7 \pm
3.3$\,pc for the final working sample (in original distance scale, see
Table~\ref{final_results}). Since the estimates were obtained by optimizing the order of
the model~$\zeta_n$, this mismatch should mainly be a consequence of a combination of the
flaring and random errors in distances. Indeed, the observed deviation of objects relative
to the model~$\zeta_{n_\text{o}}$ is due to two factors: the {\em natural\/} (true,
cosmic) dispersion of objects relative to the average surface of the Galactic disk and
the measuring dispersion---the deviation of the observed positions of objects from
their true positions due to random errors in distance moduli. The latter means that more
distant objects have larger distance errors. On the other hand, the natural dispersion can
grow to the periphery (the disk flaring). Indeed, at $X<0$\,kpc, the apparent spread of
objects relative to the model increases somewhat (Figure~\ref{Cepheids_cuts}, left
panels). In order to separate the contributions of these two effects, a more complex
version of the present method is required with direct consideration of random errors in
distances. This will allow you to get more reliable results about flaring itself.

Note that the contribution of the measuring dispersion to the observed
dispersion~$\sigma_\rho$ in this case is very small, as shown by the following approximate
estimates. For photometric distances, their standard deviation due to measuring dispersion
is $\sigma_r=\frac{\ln 10}{5}\,\sigma_d\,r$, where $\sigma_d$ is the standard uncertainty
of distance moduli. In the case of local sample, the main contribution of distance errors
to the observed vertical standard deviation~$\sigma_Z$ (close to~$\sigma_\rho$) occurs due
to objects at high latitudes $b$. Then the assumption that the standard~$\sigma_r$ for
{\em all\/} objects of the sample completely passes into vertical standard deviation gives
an upper estimate for the contribution of the measuring dispersion to~$\sigma_Z$:
$\sigma_{Z,\text{mes}}<\sigma_r$. For the distance $r=1\sigma_\rho=76.5$\,pc (see
Table~\ref{final_results}) and $\sigma_d=0.14^m$ (\citealt{Veselova+Nikiforov2020}), this
results in $\sigma_{Z,\text{mes}}<4.9$\,pc and the natural dispersion
$\sigma_{Z,0}=\sqrt{\sigma_Z^2-\sigma_{Z,\text{mes}}^2}>76.3$\,pc,  i.e., the correction
is no more than $-0.2\%$.

In the case of a working sample, most of the objects are located at small $|b|$: for
characteristic distances $r=3$--$6$\,pc (see Fig.~\ref{Cepheids_XY}) and
$Z=1\sigma_\rho=132$\,pc (Table~\ref{final_results}) $|b|=1\fdg3$--$2\fdg5$. For sample
objects, the contribution of the measuring dispersion is
$\sigma_{Z,\text{mes}}=\sigma_r\sin|b|$, $\tan b=Z/(r\cos b)$, then for small~$|b|$
$b\approx Z/r$, $\sigma_{Z,\text{mes}}\approx\sigma_r|b|=\frac{\ln 10}{5}\,\sigma_d\,|Z|$.
Then for $Z=1\sigma_\rho=132$\,pc it turns out: $\sigma_{Z,\text{mes}}\approx 8.5$\,pc,
$\sigma_{Z,0}\approx 131.7$\,pc, i.e., correction $-0.2\%$. Thus, in both cases, the
contribution of random distance uncertainty to the observed vertical dispersion is
negligible.

%, but not by $Y<0$\,kpc

From classical Cepheids on $r<2$\,kpc
\citet{Majaess+2009} obtained $z_{\sun} =26 \pm 3$\,pc and the scale height of ${\le} 75 \pm 10$\,pc
which are consistent with our estimates (Table~\ref{final_results}). Estimates of
$z_{\sun} = (23$--$24) \pm 2\;\text{pc}$ and $\sigma_{\rho} = 76.4 \pm 1.8\;\text{pc}$
were found by \citet{Bobylev+Bajkova2016_2} for classical Cepheids according to the same
version of the  \citeauthor{Berdnikov+2000}'s catalog as in this work.
\citet{Bobylev+Bajkova2016_2} considered the cylindrical region~$r\le4$\,pc. As they used
the original calibration of the catalog we can compare these estimates with ours in the
same calibration (Table~\ref{final_results}). There is consistency with our estimates of
the solar offset for both final local and final working samples. However, the vertical
scale estimates are consistent only in the case of final local sample. Exactly the same
situation is for Cepheids-based estimates in \citet{Skowron+2019}. The authors considered
data on 2431 Cepheids, and the most part of the data was obtained by the OGLE-IV project.
\citeauthor{Skowron+2019} obtained an estimate of the disk scale height of $73.5\pm3.2$ pc, so our
estimate for the final local sample does not contradict this value. All this also points
out the importance of taking into account the warp of the Galactic average disk surface in
order to have the ability of proper consideration of all the data available. Otherwise,
only local regions can be considered.

In addition, the fact that the estimates of $z_{\odot}$ differ, as noted in Introduction,
also suggests that the a priori assumption about the flat model of the Galactic disk
should be limited. Indeed, according to our results such assumption might be made only for
specific regions like the local one, i.e., close enough to the Sun. Moreover, it can be
noticed that the value of $z_{\sun}$ is less dependent on the Galactic disk warping than
the value of $\sigma_{\rho}$. Based on what has been said, we can conclude that any a
priori assumption about the Galactic disk warping must be carefully studied especially
when the vertical scale parameter is estimated.

The detected local extrema of the average surface of the disk may be manifestations of
bending waves caused by interaction with the Sagittarius dwarf galaxy, in the form of
local structures elongated in the azi\-mu\-thal direction (\citealt{Gomez+2013}, fig.~5;
\citealt{Laporte+2019}; \citealt{Poggio+2021}, fig.~2), or by interaction with the Large
Magellanic Cloud (\citealt{Thulasidharan+2022} and references therein).

The method used after testing on classical Cepheids can now be applied to other data (in
particular, to Gaia data) and/or in other assumptions about the distribution function
$f(\rho)$. The Gaia DR2 catalog was used in recent work by \citet{Ablimit_2020} to obtain
data on classical Cepheids. Despite the fact the direct study of the Galactic disk warping
was not conducted in the work of these authors, according to the pictures plotted in the
work on these data, the disk warping is clearly revealed. Unfortunately,
the use of the current version of our method with these data as is will lead to
significant biases mainly because of  the dependence of distance uncertainty on distance,
which will be significant due to the need to consider the large neighborhood of the Sun.

% to other types of reference objects

%So the distance errors must be taken into account in order to use Gaia DR2.

Taking into account the uncertainty of distances may also solve the problem of
establishing the form of the vertical distribution law $f(\rho)$. Note that the analysis
of the 2D distribution does not allow us to draw a definite conclusion about the
functional form of this law (\citealt{Mosenkov+2021}).

In the future, we plan to apply the proposed method in the variant of accounting for
distance uncertainties to new databases.

%Overall, we can conclude the method can already be applied as is for modern data.
%Moreover, obtained results reveal the new information about the structure of Galactic disk
%surface warping. However,  taking into account the distance errors is naturally important
%for data obtained using parallax observations due to the different accuracy of large and
%small distances, so that step is strongly recommended for such data. In our future work we
%plan to include an accounting of errors of the distances in the proposed algorithm.

% section conclusion (end)

% \normalem
\begin{acknowledgements}

We are grateful for the helpful comments from the
anonymous referee.

\end{acknowledgements}
  
\bibliographystyle{raa}
\bibliography{bibtex}

\begin{thebibliography}{99}
\providecommand\natexlab[1]{#1}
\providecommand\JournalTitle[1]{#1}

\bibitem[Ablimit {et~al.}(2020)]{Ablimit_2020}
Ablimit, I., Zhao, G., Flynn, C.,  Bird, S.~A. 2020, \apj, 895, L12

\bibitem[Berdnikov {et~al.}(2000)]{Berdnikov+2000}
Berdnikov, L., Dambis, A.,   Vozyakova, O. 2000, Astron. Astrophys. Suppl.
  Ser., 143, 211

\bibitem[Binney \& Merrifield(1998)]{BM98} 
Binney J., \& Merrifield M. 1998, Galactic Astronomy (Princeton, USA:
Princeton Univ. Press)

\bibitem[Bland-Hawthorn \& Gerhard(2016)]{Bland-Hawthorn+2016}
Bland-Hawthorn, J., \& Gerhard, O. 2016, Annual Rev. Astron. Astrophys., 54,
  529

\bibitem[Bobylev \& Bajkova(2016{\natexlab{a}})]{Bobylev+Bajkova2016_2}
Bobylev, V., \& Bajkova, A. 2016{\natexlab{a}}, Astron. Let., 42, 1

\bibitem[Bobylev \& Bajkova(2016{\natexlab{b}})]{Bobylev+Bajkova2016_1}
Bobylev, V., \& Bajkova, A. 2016{\natexlab{b}}, Astron. Let., 42, 182.

\bibitem[Bobylev \& Bajkova(2017)]{Bobylev+Bajkova2017}
Bobylev, V., \& Bajkova, A. 2017, Astron. Let., 43, 304.

\bibitem[Burke(1957)]{Burke1957}
Burke, B. F. 1957, \aj, 62, 90.

\bibitem[Buckner \& Froebrich(2014)]{Buckner+Froebrich2014}
Buckner, A. S.~M., \& Froebrich, D. 2014, \mnras, 444, 290.

%\bibitem[Carballo-Bello {et~al.}(2021)]{Carballo-Bello+2021}

\bibitem[Cheng {et~al.}(2020)]{Cheng+2020}
 Cheng, X., Anguiano, B., Majewski, S.R., et al.
%Hayes, C., Arras, P., Chiappini, C.,
%Hasselquist, S., Queiroz, A. B. A., Nitschelm, C., Anıbal Garcıa-Hern\'andez, D.,
%Lane, R. R., Roman-Lopes, A., Frinchaboy, P. 
2020, ApJ, 905, 49.
%arXiv e-prints, arXiv:2010.10398

\bibitem[Chrob\'akov\'a \& L\'opez-Corredoira(2021)]{Chrobakova+LC2021}
Chrob\'akov\'a, \v{Z}., \& L\'opez-Corredoira, M. 2021, ApJ, 912, 130. %arXiv e-prints, arXiv:2105.04348.

\bibitem[Chrob\'akov\'a {et~al.}(2020)]{Chrobakova+2020}
Chrob\'akov\'a, \v{Z}., Nagy, R., \& L\'opez-Corredoira, M. 2020,  \aap, 637, 96.

\bibitem[Dambis {et~al.}(2015)]{Dambis+2015}
Dambis, A. K., Berdnikov, L. N., Efremov, Yu. N., et al.
2015, Astron. Lett., 41, 489.

\bibitem[Efremov(2011)]{Efremov2011}
Efremov, Yu. N. 2011, Astronomy Reports, 55, 108.  %  Astronomy Reports, Volume 55, Issue 2, pp.108-122

\bibitem[de~Grijs {et~al.}(2014)]{deGrijs+2014}
de~Grijs, R., Wicker, J., \& Bono, G. 2014, \aj, 147, 122.

\bibitem[Ferguson {et~al.}(2017)]{Ferguson+2017}
Ferguson, D., Gardner, S., \& Yanny, B. 2017, ApJ, 843, 141.

%\bibitem[Garavito-Camargo {et~al.}(2019)]{Garavito-Camargo+2019}

\bibitem[G\'{o}mez {et~al.}(2013)]{Gomez+2013}
G\'{o}mez, F. A., Minchev, I., O'Shea, B. W., et al. 2013, \mnras, 429, 159.

\bibitem[Hudson(1964)]{Hudson1964}
Hudson, D. J. 1964, Statistics: Lectures on Elementary Statistics and Probability (Geneva: CERN)

\bibitem[Juri\'c {et~al.}(2008)]{Juric+2008}
Juri\'c, M., Ivezi\'c, \v{Z}., Brooks, A., Lupton, R., \& Schlegel, D. 2008, ApJ,
  673, 864.

\bibitem[Kerr(1957)]{Kerr1957}
Kerr, F. J. 1957, \aj, 62, 93.

\bibitem[Khachaturyants {et~al.}(2021)]{Khachaturyants+2021}
Khachaturyants, T., Beraldo e Silva, L., \& Debattista, V. P. 2021, \mnras, 508, 2350.

\bibitem[Khachaturyants {et~al.}(2022)]{Khachaturyants+2022}
Khachaturyants, T., Beraldo e Silva, L., Debattista, V. P., \& Daniel, K. J. 2022, \mnras, 512, 3500.

\bibitem[Laporte {et~al.}(2019)]{Laporte+2019}
Laporte, C. F. P., Minchev, I., Johnston, K. V., \&  G\'omez, F. A. 2019, \mnras, 485, 3134.

\bibitem[Lemasle et al. (2022)]{Lemasle+2022}
Lemasle, B., Lala, H.N., Kovtyukh, V., {et~al.} 2022, A\&A, 668, 40.
%e-print arXiv:2209.02731

\bibitem[Li {et~al.}(2020)]{Li+2020} 
Li, X.-Y., Huang, Y., Chen, B.-Q., et al.
% Wang, H.-F., Sun, W.-X., Guo, H.-L., Li, Q.-Z., Liu, X.-W.
2020, ApJ, 901, 56.

\bibitem[Majaess {et~al.}(2009)]{Majaess+2009}
Majaess, D. J., Turner, D. G., \& Lane D. J. 2009, \mnras, 398, 263.

\bibitem[Mel'nik {et~al.}(2015)]{Mel'nik+2015}
Mel'nik, A., Rautiainen, P., Berdnikov, L., Dambis, A., \&  Rastorguev, A. 2015,
Astron. Nachr., 336, 70.

\bibitem[Minniti {et~al.}(2021)]{Minniti+2021}
Minniti, J. H., Zoccali, M., Rojas-Arriagada, A., et al.
%Minniti, D., Sbordone, L., Contreras Ramos, R., Braga, V.F.,; Catelan, M., Duffau, S., Gieren, W., Marconi, M., Valcarce, A.A.R.
2021, \aap, 654, A138.

\bibitem[Mosenkov {et~al.}(2021)]{Mosenkov+2021}
Mosenkov, A. V., Savchenko, S. S., Smirnov, A. A., \& Camps, P. 2021, \mnras, 507, 5246.

\bibitem[Nikiforov(2012)]{Nikiforov2012}
Nikiforov, I. I. 2012, Astron. Astrophys. Trans., 27, 537.

\bibitem[Nikiforov \& Veselova(2018)]{Nikiforov+Veselova2018a}
Nikiforov, I. I., \& Veselova, A. V. 2018, Astronomy Letters, 44, 81.

\bibitem[Oort {et~al.}(1958)]{Oort+1958}
Oort, J., Kerr, F., \&  Westerhout, G. 1958, \mnras, 118, 379.

\bibitem[Poggio {et~al.}(2020)]{Poggio+2020}
Poggio, E., Drimmel, R., Andrae, R., et al.
%Bailer-Jones, C.A.L., Fouesneau, M., Lattanzi, M.G.,
%Smart, R.L.,   Spagna, A.
2020, NatAst, 4, 590.

\bibitem[Poggio {et~al.}(2021)]{Poggio+2021}
Poggio, E.,  Laporte, C. F. P., Johnston, K. V., et al.
%D'Onghia, E., Drimmel, R.,  Grion Filho, D.
2021, \mnras, 508, 541.

\bibitem[Pohl {et~al.}(2008)]{Pohl+2008}
Pohl, M., Englmaier, P., \& Bissantz, N. 2008, \aj, 677, 283.

\bibitem[Pojmanski (2002)]{Pojmanski2002}
Pojmanski, G. 2002, Acta Astron., 52, 397.

\bibitem[Reid {et~al.}(2019)]{Reid+2019}
Reid, M.J., Menten, K.M., Brunthaler, A., et al.
%Zheng, X.W., Dame, T.M., Xu, Y., Li, J., Sakai, N., Wu, Y., Immer, K., Zhang, B., Sanna, A., Moscadelli, L., Rygl, K.L.J., Bartkiewicz, A., Hu, B.,
%Quiroga-Nu\~{n}ez, L.H., van Langevelde, H.J.
2019, \aj, 885, 131.

\bibitem[Romero-G\'omez {et~al.}(2019)]{Romero-Gomez+2019}
Romero-G\'omez, M., Mateu, C., Aguilar, L., Figueras, F., \&  Castro-Ginard, A. 2019, \aap, 627, A150.

\bibitem[Samus {et~al.}(2017)]{Samus+2017}
Samus, N. N., Kazarovets, E. V., Durlevich, O. V., et al. 2017, ARep, 61, 80.
%\bibitem[Samus {et~al.}(2007--2017)]{Samus+2007-2017}
%Kireeva N.N., Pastukhova E.N.
%VizieR On-line Data Catalog, 
%General Catalogue of Variable Stars: B/gcvs (2007--2017).

\bibitem[Schoenrich \& Dehnen(2018)]{Schoenrich+Dehnen2018}
Sch\"onrich, R., \& Dehnen, W. 2018, \mnras, 478, 3809.

\bibitem[Skowron {et~al.}(2019a)]{Skowron+2019}
Skowron, D., Skowron, J., Mr\'oz, P., {et~al.} 2019a, Science, 365, 478.

\bibitem[Skowron {et~al.}(2019b)]{Skowron+2019b}
Skowron, D., Skowron, J., Mr\'oz, P., {et~al.} 2019b, Acta Astronomica, 69, 305.

\bibitem[Spicker \& Feitzinger(1986)]{Spicker+Feitzinger1986}
Spicker, J., \& Feitzinger, J. V. 1986, \aap, 163, 43.

\bibitem[Thulasidharan {et~al.}(2022)]{Thulasidharan+2022}
Thulasidharan, L., D'Onghia, E., Poggio, E., et al. 2022, \aap, 660, L12.

%\bibitem[van Leeuwen (2007)]{van+Leeuwen2007}
%van Leeuwen, F. 2007, \aap, 474, 653.

\bibitem[van Tulder(1942)]{vanTulder1942}
van Tulder, J. 1942, Bull. Astron. Netherlands, 9, 315.

\bibitem[Veselova \& Nikiforov(2020)]{Veselova+Nikiforov2020}
Veselova, A. V., \& Nikiforov, I. I. 2020, \raa, 20, 209.

\bibitem[Wall \& Jenkins(2012)]{Wall+2012}
Wall, J., \& Jenkins, C. 2012, Practical statistics for astronomers. Second
Edition (New York, USA: Cambridge University Press).

\bibitem[Xu {et~al.}(2015)]{Xu+2015}
Xu, Y., Newberg, H. J., Carlin, J. L., et al. 2015, \apj, 801, 105. 

\bibitem[Yao {et~al.}(2017)]{Yao+2017}
Yao, J., Manchester, R., \&  Wang, N. 2017, \mnras, 468, 3289.

%\bibitem[Yu {et~al.}(2021)]{Yu+2021}
%Yu, Y., Wang, H.-F., Cui, W.-Y.,  et al.
% Li, L.-L., Liu, C., Zhang, B., Tian, H., Huo, Z.-Y., Ju, J., Liu, Z.-C., Wen, F., Feng, Sh.
%2021,  ApJ, 922, 80. 
%arXiv e-prints, arXiv::2102.00731.

\end{thebibliography}

\end{document}